\documentclass[journal, 12pt, draftclsnofoot, onecolumn]{IEEEtran}
\IEEEoverridecommandlockouts

\usepackage{amsfonts,amssymb}
\usepackage{ifpdf}
\usepackage{cite}
\usepackage{stfloats}
\usepackage{bm}
\usepackage{color,xcolor}
\usepackage{subfigure}
\ifCLASSINFOpdf
   \usepackage[pdftex]{graphicx}
   \graphicspath{{../pdf/}{../jpeg/}}
   \DeclareGraphicsExtensions{.pdf,.jpeg,.png}
\else
   \usepackage[dvips]{graphicx}

   \graphicspath{{../eps/}}

   \DeclareGraphicsExtensions{.eps}
\fi

\usepackage[cmex10]{amsmath}

\usepackage{algorithm}
\usepackage{algorithmic}
\ifCLASSOPTIONcompsoc
\else
  \usepackage[caption=false,font=footnotesize]{subfig}
\fi

\usepackage{makecell}
\begin{document}
	\title{A Novel Wireless Communication Paradigm for Intelligent Reflecting Surface Based  Symbiotic Radio Systems}
	
\author{Meng~Hua,
Qingqing~Wu,~\IEEEmembership{Member,~IEEE,}
 Luxi~Yang,~\IEEEmembership{Senior Member,~IEEE,}
  Robert~Schober,~\IEEEmembership{Fellow,~IEEE,}
  H.~Vincent~Poor,~\IEEEmembership{Life Fellow,~IEEE}
%

\thanks{M. Hua and Q. Wu are with the State Key Laboratory of Internet of Things for Smart City and Department of Electrical and Computer Engineering, University of Macau, Macao 999078, China (email: menghua@um.edu.mo; qingqingwu@um.edu.mo). }
\thanks{ L. Yang is with the School of Information Science and Engineering, Southeast University, Nanjing 210096, China (e-mail: lxyang@seu.edu.cn).}
\thanks{R. Schober are with the Institute for Digital Communications, Friedrich-Alexander-University Erlangen-Nurnberg, Erlangen 91054, Germany (e-mail: robert.schober@fau.de).}
\thanks{H. V. Poor is with the Department of Electrical Engineering, Princeton
	University, Princeton, NJ 08544, USA (e-mail: poor@princeton.edu).}

}
\maketitle
\vspace{-3em}
\begin{abstract}
This paper investigates a novel  intelligent reflecting surface (IRS)-based symbiotic radio (SR) system architecture consisting  of a transmitter, an IRS, and   an  information receiver (IR). The primary transmitter communicates with the IR and at the same time assists the IRS in forwarding information to the IR.
 Based on the IRS's  symbol period,  we distinguish  two scenarios, namely,    commensal  SR (CSR) and parasitic SR (PSR), where two different  techniques   for   decoding the IRS signals at the IR are employed. We formulate  bit error rate (BER) minimization problems for both scenarios by jointly optimizing the active beamformer at the  base station and  the  phase shifts at the IRS, subject to a minimum primary rate requirement.  Specifically, for the  CSR scenario, a  penalty-based algorithm is proposed to obtain a high-quality solution,  where semi-closed-form solutions for the active beamformer and the IRS phase shifts are derived  based on   Lagrange duality  and Majorization-Minimization methods, respectively. For the PSR scenario,  we apply a bisection search-based method, successive convex approximation, and difference of convex  programming  to develop a computationally efficient algorithm, which converges to   a locally optimal solution.
 Simulation results demonstrate the effectiveness of  the proposed  algorithms  and show that  the  proposed SR techniques   are able to achieve a lower BER than    benchmark schemes.
\end{abstract}


\section{Introduction}
Recently, intelligent reflecting surfaces (IRSs), also termed reconfigurable intelligent surfaces (RISs), have  attracted significant  attention  from both academia and industry \cite{marco2020smart,wu2020intelligentarxiv,huang2020Holographic}. IRSs are composed of  large numbers  of reflecting elements  (e.g., low-cost printed dipoles) \cite{WU2020towards}. The  elements  of an  IRS  are based on   metamaterials with   subwavelength structure and  are able to adjust the incident signal's amplitude, phase, frequency, and 	polarization,  thus being able to  collaboratively
change the reflected signal's propagation \cite{cui2017information}. Different from traditional reflecting surfaces, where   the phase shift is fixed after   fabrication,  the phase shifters of IRSs can be  dynamically adjusted   between  $0$ and $2\pi$  to adapt to  varying wireless  channel conditions \cite{Zhu2013Active}. In addition, different from current base stations  (BSs)/active relays, which require  power-hungry and high-cost radio frequency (RF) chains,  	IRSs are much    greener and more cost-effective due to their simple integrated passive components, such  as  varactor diodes,  positive-intrinsic-negative (PIN) diodes, micro-electro-mechanical system
(MEMS) switches, and field-effect transistors (FETs)\cite{wu2020intelligentarxiv}. Furthermore, 
IRSs can be fabricated as artificial thin films and  readily  attached to existing infrastructures, such as the facades of buildings, indoor ceilings, and even smart t-shirts \cite{marco2020smart}, thus making them  promising for  implementation  in practice.  Due to the  above appealing benefits,  IRSs have been recognized as a key  solution for improving both the spectral and energy efficiency in future sixth-generation (6G)  cellular wireless networks.

By  properly adjusting the  phase shifts of a large number of IRS reflecting elements,  the signals reflected  by a planar IRS   coherently add up at   desired receivers to boost  the received power, while they add up destructively   at  non-intended receivers to suppress   co-channel interference \cite{wu2019intelligent,beamforming2020wu}. For example, it was shown in   \cite{wu2019intelligent}   that  the received signal-to-noise ratio (SNR) increases quadratically with the number of reflecting elements in  a single-user IRS-aided
system, which    unveiled the fundamental scaling law of IRS. Subsequently, various follow-up 
 works have investigated  the application of IRSs for other purposes,
 such as  physical layer security \cite{Intelligent2020guan,zhang2020robust,yu2020robust,feng2020large}, multi-cell cooperation \cite{pan2020multicell,hua2020intelligent,xie2020max}, simultaneous wireless information and
power transfer \cite{wu2019weighted,pan2019intelligent,li2020joint}, and unmanned aerial vehicle communication  \cite{pan2020uav,hui2019Reflections,sixian2020robust}. Due to the similarities  between IRSs and active relaying, some  works compared the performance gain provided  by IRSs with that of relays
\cite{Boulogeorgos2020Performance},\cite{bjornson2019intelligent}. In  \cite{Boulogeorgos2020Performance}, the authors studied the classical three-node cooperative transmission system  and compared   the performances of  IRSs with that   of amplify-and-forward (AF) relays. The results showed that IRS-assisted wireless systems outperform   AF relaying wireless systems  in terms of the average SNR,
outage probability, symbol error rate, and ergodic capacity when the aperture  of the IRS is sufficiently  large. Similar results were also obtained in \cite{bjornson2019intelligent} for the comparison of   IRSs and   decode-and-forward (DF) relays with respect to the maximum energy efficiency and the required total transmit power.


Different from  the above studies,  where   IRSs were used only  to assist the communication  of  
existing communication  systems, a new IRS functionality referred to as symbiotic radio (SR) transmission (also known  as passive beamforming and information transfer transmission) was
proposed recently \cite{yan2020passive,yan2020wlcpassive,hu2020reconfigurable,lin2020Reconfigurable,zhang2020large,hua2020uav}.  The preliminary concept of simultaneous  passive beamforming and information transfer   was introduced  in \cite{yan2020passive,yan2020wlcpassive,hu2020reconfigurable},  where  the IRS did  not only  help the transmitter enhance the transmission of the  primary wireless network  via  passive beamforming  but  also   delivered its own information to 
 receivers by leveraging the reflected signals. For example, a sensing device, which is  able to sense and collect   environmental information such as  illuminating light, temperature, and humidity,
 can be  connected to the smart  controller of an  IRS,  and  the smart controller conveys the sensed information (i.e., a sequence of  $0$ and   $1$ symbols) to the desired receiver via adjusting  the  on/off state of the  IRS. Then, the receiver decodes the information based on the   differences of the responses of the IRS for  the two states. As such, the information is encoded into the on/off state of the IRS. This concept is similar to spatial modulation transmission, where the indices of active transmit antennas are  exploited to encode  information to improve    spectral efficiency \cite{Mesleh2008Spatial}.

In this paper, we study a novel wireless communication paradigm for IRS-based SR systems. We consider  a network  consisting of a BS, an IRS, and an information receiver (IR).  The BS and  the IR constitute  the primary network, and  the BS transmits the primary  information to the IR.  The IRS is deployed nearby the  IR, and leverages the radio wave generated by the BS to deliver its own information to the  IR by 
 adjusting  its  on/off state.
We aim at minimizing the bit error rate (BER) of the IRS  by jointly optimizing the BS beamformer  and    IRS phase shifts while guaranteeing a minimum  required rate for  the primary network  subject to the BS transmit power budget and   the unit-modulus phase-shift constraints.  Based on   the  IRS's   symbol period,  two scenarios, namely,    commensal  SR (CSR) and parasitic SR (PSR), are considered. For the CSR scenario, 
the IRS's  symbol period is much smaller than that of  the primary transmission. During one primary symbol period, the IRS's transmission can be regarded as  an additional multipath component for  the primary transmission.  As such, the IRS is also used to strengthen the primary network's  transmission.
 In contrast, for the  PSR scenario, where the IRS's  symbol period is comparable to  that of the primary transmission,  the IRS's signal is treated as interference when decoding the primary symbol at the receiver. Therefore,   different decoding techniques are needed for the above two  scenarios, which leads to    different expressions for the   objective function. We note    that the  proposed IRS based SR is significantly different from    backscatter based  SR \cite{long2019symbiotic},\cite{long2019full}. Specifically, in the considered system, 
the IRS  acts  not only  as an information source node but also as a helper 
 for  improving the performance of the primary link
via passive beamforming. In contrast,     backscatter tags are only information sources that transmit  their
own signals to the receiver by riding on the sinusoidal signal generated by the transmitter. 
As such, the transmission model and   the    problem formulation  in our paper differ   from that for  backscatter based  SR. Furthermore,
compared to \cite{yan2020passive,yan2020wlcpassive,hu2020reconfigurable,lin2020Reconfigurable,zhang2020large,hua2020uav}, this paper is the first work  that targets the   minimization of the BER of the IRS  symbols while taking account into the primary rate requirements.  We propose a novel algorithm, namely, the  penalty-based algorithm, to solve this problem. In addition,  this is also the first work to consider     PSR for  IRSs, and propose a corresponding bisection  search-based  algorithm. 
The main contributions of this paper are summarized as follows:
\begin{itemize}
	\item For the CSR scenario,  we formulate an optimization problem for minimization of the BER, which is shown to be  non-convex. To solve this problem efficiently, a novel penalty-based algorithm is proposed, which comprises  a two-layer iteration, i.e.,  an inner layer iteration and an outer  layer iteration. The inner layer solves the penalized optimization problem, while the  outer layer updates the penalty coefficient from one iteration to the next to guarantee convergence.  In particular,  in the inner layer,   semi-closed-form solutions for both the   BS beamformer and  the  IRS phase shifts are obtained based on  Lagrange duality   and Majorization-Minimization (MM) techniques, respectively.
	\item For the PSR scenario,   the formulated BER minimization problem  is rather  complicated  and fundamentally different from that for  CSR. To overcome this difficulty, a bisection search-based algorithm is proposed.  We first derive  the  search range space, and then find the desired  point by checking the  feasibility of the resulting  problem. To reduce the computational complexity,  a semi-closed-form expression  for the BS beamformer is derived. To incorporate the unit-modulus IRS phase-shift constraint,  we leverage the difference of convex (DC) programming framework  instead of the commonly used  semidefinite
	relaxation  (SDR) technique to avoid the high probability of  obtaining    non-rank-one solutions. 
	\item Simulation results demonstrate that  for  both   scenarios, i.e.,   CSR and PSR,  the proposed algorithms for   joint BS beamformer
	and IRS phase shift  optimization   outperform   benchmark schemes employing     maximum ratio transmission (MRT)   and random IRS  phase shifts, respectively. We also find that  the BER of   CSR  is much smaller than that of PSR, since    the interference  can be  harnessed in the former case.	Furthermore, we unveil that  deploying IRSs nearby the BS or the IR  can significantly improve  the system performance for both scenarios.
\end{itemize}
The rest of this paper is organized as follows. Section II introduces the system model and problem formulation  for the  CSR and PSR scenarios, respectively. In Sections III and IV, we propose efficient algorithms for the two resulting  optimization  problems, respectively. Numerical results are provided in Section V, and  the paper is concluded in Section VI.

\emph{Notations}: Boldface lower-case and upper-case letters denote  vectors and matrices, respectively.  ${\mathbb C}^ {d_1\times d_2}$ stands for the set of  complex $d_1\times d_2$  matrices. For a complex-valued vector $\bf x$, ${\left\| {\bf x} \right\|}$ represents the  Euclidean norm of $\bf x$, ${\rm arg}({\bf x})$ denotes  the phase of   $\bf x$,  and ${\rm diag}(\bf x) $ denotes a diagonal matrix whose main diagonal elements are extracted from vector $\bf x$.
 For a square matrix $\bf X$, ${\bf X}^*$, ${\bf X}^H$, ${\rm{Tr}}\left( {\bf{X}} \right)$,  ${{\bf{X}}^{ - 1}}$, ${ {\bf{X}} ^\dag }$,   ${\rm{rank}}\left( {\bf{X}} \right)$, and ${\left\| {\bf{X}} \right\|_2}$ stand for  its conjugate, conjugate transpose, trace,  inverse, pseudoinverse, rank, and $l$-2 norm, respectively.  ${\bf{X}} \succeq {\bf{0}}$ indicates that matrix $\bf X$ is a positive semi-definite matrix. ${\left[ {\bf{X}} \right]_{i,i}}$ represents the $i$th main diagonal element of   matrix $\bf X$. ${\bf I}$ and $\bf 0$ denote the  identity matrix and all-zeros matrix with appropriate dimensions, respectively.
 A circularly symmetric complex Gaussian (CSCG) random variable $x$ with mean $ \mu$ and variance  $ \sigma^2$ is denoted by ${x} \sim {\cal CN}\left( {{{\mu }},{{\sigma^2 }}} \right)$. A real Gaussian random variable $x$ with mean $ \mu$ and variance  $ \sigma^2$ is denoted by ${x} \sim {\cal N}\left( {{{\mu }},{{\sigma^2 }}} \right)$.  Statistical expectation and statistical variance  are denoted by ${\mathbb E}\left\{  \cdot  \right\}$ and  ${\mathbb V {\rm ar}}\left\{  \cdot  \right\}$, respectively. ${\mathop{\rm Re}\nolimits} \left\{ {x} \right\}$ denotes the  real part of a complex variable $ x$.  ${\cal O}\left(  \cdot  \right)$ is the big-O computational complexity notation.

\begin{figure}[!t]
\centerline{\includegraphics[width=2.8in]{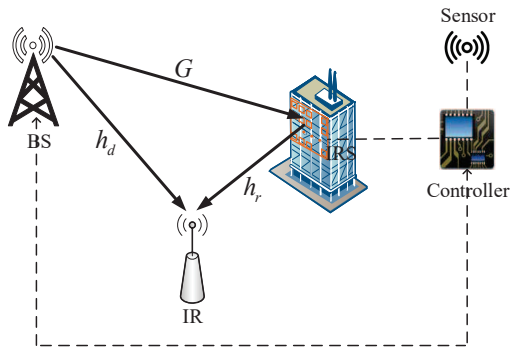}}
\caption{A novel wireless communication paradigm for IRS based  SR systems.} \label{fig1}
	\vspace{-20pt}
\end{figure}
\section{System Model and Problem Formulation}
\subsection{System Model}
As shown in Fig.~\ref{fig1}, we consider an IRS-based  SR system consisting  of a   BS, an IRS, and an IR, where the BS transmits   primary signals to the IR  and the IRS  delivers its own information to the IR by leveraging  radio waves generated by the BS. We assume that the   BS is equipped with $N$ transmit antennas, the IR is equipped with one  antenna, and the IRS has $M$ reflecting elements.  Let ${\bf{h}}_{d} \in {{\mathbb  C}^{N \times 1}}$, ${\bf G} \in {{\mathbb C}^{M \times {N}}}$, and ${\bf h}_{r} \in {{\mathbb C}^{M \times 1}}$ denote the complex equivalent baseband  channels  between the BS  and  the  IR, between the  BS  and the IRS, and  between  the  IRS and the IR, respectively. The IRS reflection    can be characterized by a diagonal reflection coefficient  matrix  ${{\bf{\Theta }}}{\rm{ = diag}}\left( {{e^{j{\theta _{1}}}}, \ldots ,{e^{j{\theta _{M}}}}} \right)$, where  the reflection amplitude is  fixed as   $1$,  and  $\theta_{m}$ denotes  the   phase shift corresponding to the $m$th  IRS reflecting element\cite{WU2020towards}, \cite{wu2019intelligent},\cite{zhang2020Capacity}.

The CSR   and PSR scenarios considered in this paper  are described in the following.
\subsubsection{CSR scenario} In the CSR scenario, the symbol rate of  the IRS transmission is much smaller than that of  the primary
 transmission due to the limited computational and communication capabilities at the IRS. Denote   the durations of  the  IRS symbol and the primary symbol by $T_s$ and $T_x$, respectively. Without loss of generality, we assume that  each IRS symbol spans $ L$ primary symbols, i.e., $T_s= LT_x$.  Denote by  $x[l]$, $0 \le l \le L$, and $s$ the BS's $l$th symbol  and the IRS's  symbol, which is generated by  the  on/off state of the IRS, respectively. The  $l$th symbol  received by the IR  is given by 
 \begin{align}
 {y_{csr,r}}\left[ l \right] = \underbrace {{\bf{h}}_d^H{\bf{w}}x\left[ l \right]}_{{\rm{direct}}{\kern 1pt} {\kern 1pt} {\rm{link}}} + \underbrace {{\bf{h}}_r^H\left( {s{\bf{\Theta }}} \right){\bf{Gw}}x\left[ l \right]}_{{\rm{reflected}}{\kern 1pt} {\kern 1pt} {\rm{link}}} + {n_r[l]}, \label{csr_combinatedsignal}
 \end{align}
 where ${\bf{w}} \in {{\mathbb  C}^{N \times 1}}$ is the transmit beamforming vector at the BS, ${x[l]} \sim {\cal CN}\left( {0,1} \right)$, and ${n_r[l]} \sim {\cal CN}\left( {0,{\sigma ^2}} \right)$ denotes the additive white Gaussian noise  at the IR. We adopt the simple but widely used on-off keying
 (OOK) modulation for the  information transmission of the IRS, i.e., $s = \left\{ {0,1} \right\}$.
 We assume that the probability for the IRS to send symbol ``1''  is $\rho$, and that to  send symbol ``0''  is $1-\rho$. Without loss of generality, we assume the IRS sends  symbol ``1'' and symbol ``0'' with equal probability, i.e., $\rho=\frac{1}{2}$. We note that symbol ``1'', i.e., $s=1$, implies  the IRS is turned on and  symbol ``0'', i.e., $s=0$, implies  the IRS is turned off.

 The instantaneous  achievable rate (bps/Hz) of the IRS-assisted primary system   is given by
 \begin{align}
 {\tilde R_{csr,x}}\left( s \right) = {\log _2}\left( {1 + \frac{{{{\left| {{\bf{h}}_d^H{\bf{w}} + {\bf{h}}_r^H\left( {s{\bf{\Theta }}} \right){\bf{Gw}}} \right|}^2}}}{{{\sigma ^2}}}} \right).
 \end{align}
Since the instantaneous achievable rate ${\tilde R_{csr,x}}\left( s \right)$ depends on the IRS's on/off state,  the average 
achievable rate of  the primary system is  given by \cite{zhang2020large},\cite{long2019symbiotic}
\begin{align}
{R_{csr,x}} &= {\mathbb E}_s\left\{ {{{\tilde R}_{csr,x}}\left( s \right)} \right\}\notag\\
&= \left( {1 - \rho } \right){\log _2}\left( {1 + \frac{{{{\left| {{\bf{h}}_d^H{\bf{w}}} \right|}^2}}}{{{\sigma ^2}}}} \right) + \rho {\log _2}\left( {1 + \frac{{{{\left| {{\bf{h}}_d^H{\bf{w}} + {{\bf{v}}^H}{\rm{diag}}\left( {{\bf{h}}_r^H} \right){\bf{Gw}}} \right|}^2}}}{{{\sigma ^2}}}} \right).
\end{align}
where ${{\bf{v}}^H} = \left( {{e^{j{\theta _1}}}, \ldots ,{e^{j{\theta _M}}}} \right)$.

After successfully decoding the primary signal $x[l]$,   the receiver can apply  successive interference cancellation (SIC)       to   remove   ${\bf{h}}_d^H{\bf{w}}x\left[ l \right]$  from the received  composite signal in \eqref{csr_combinatedsignal}. Thus, after removing this term, we   obtain the intermediate IRS signal  as follows
\begin{align}
{{\bar y}_{csr,r}}\left[ l \right] = {{\bf{v}}^H}{\rm{diag}}\left( {{\bf{h}}_r^H} \right){\bf{Gw}}sx\left[ l \right] + {n_r[l]},
\end{align}
Since each IRS symbol spans $ L$ primary symbols for the CSR scenario, the IRS is affected by   time-selective fading.  By applying   maximal-ratio-combining (MRC), the decision can be  based on the  following  real sufficient statistic  \cite{tse2005fundamentals}
 \begin{align}
 	{{\bar y}_{csr,r}}& = {\mathop{\rm Re}\nolimits} \left\{ {\sum\limits_{l = 1}^L {{{\left( {{{\bf{v}}^H}{\rm{diag}}\left( {{\bf{h}}_r^H} \right){\bf{Gw}}x\left[ l \right]} \right)}^*}{{\bar y}_{csr,r}}\left[ l \right]} } \right\}\notag\\
 	& = {\left| {{{\bf{v}}^H}{\rm{diag}}\left( {{\bf{h}}_r^H} \right){\bf{Gw}}} \right|^2}\sum\limits_{l = 1}^L {{{\left| {x\left[ l \right]} \right|}^2}s}  + {{\bar n}_r}, \label{receiver_siganl}
 \end{align}
 where ${{\bar n}_r} = {\mathop{\rm Re}\nolimits} \left\{ {{{\left( {{{\bf{v}}^H}{\rm{diag}}\left( {{\bf{h}}_r^H} \right){\bf{Gw}}} \right)}^*}\sum\limits_{l = 1}^L {{x^*}\left[ l \right]{n_r[l]}} } \right\}$. It is not difficult to see  that 
 
 \noindent ${{{\left( {{{\bf{v}}^H}{\rm{diag}}\left( {{\bf{h}}_r^H} \right){\bf{Gw}}} \right)}^*}\sum\limits_{l = 1}^L {{x^*}\left[ l \right]{n_r}[l]} }$ is still a CSCG random variable with the  expectation and variance  given by 
 \begin{align}
 &	{\mathbb E}\left\{ {{{\left( {{{\bf{v}}^H}{\rm{diag}}\left( {{\bf{h}}_r^H} \right){\bf{Gw}}} \right)}^*}\sum\limits_{l = 1}^L {{x^*}\left[ l \right]{n_r[l]}} } \right\} = 0,	\notag\\
 &{\mathop{{\mathbb V}{\rm ar}}} \left\{ {{{\left( {{{\bf{v}}^H}{\rm{diag}}\left( {{\bf{h}}_r^H} \right){\bf{Gw}}} \right)}^*}\sum\limits_{l = 1}^L {{x^*}\left[ l \right]{n_r[l]}} } \right\} = {{{\left| {{{\bf{v}}^H}{\rm{diag}}\left( {{\bf{h}}_r^H} \right){\bf{Gw}}} \right|}^2}\sum\limits_{l = 1}^L {{{\left| {x\left[ l \right]} \right|}^2}} {\sigma ^2}}.
 \end{align}
As such, ${{\bar n}_r}$ is a real Gaussian random variable and it follows that
\begin{align}
{{\bar n}_r} \sim {\cal N}\left( {0,{{\left| {{{\bf{v}}^H}{\rm{diag}}\left( {{\bf{h}}_r^H} \right){\bf{Gw}}} \right|}^2}\sum\limits_{l = 1}^L {{{\left| {x\left[ l \right]} \right|}^2}} {\sigma ^2}/2} \right).
\end{align}
 We can rewrite \eqref{receiver_siganl} as follows
\begin{align}
{{\bar y}_{csr,r}} = \left\{ \begin{array}{l}
{\left| {{{\bf{v}}^H}{\rm{diag}}\left( {{\bf{h}}_r^H} \right){\bf{Gw}}} \right|^2}\sum\limits_{l = 1}^L {{{\left| {x\left[ l \right]} \right|}^2}}  + {{\bar n}_r},{\kern 1pt} {\kern 1pt} {\kern 1pt} s = 1\\
{{\bar n}_r},{\kern 1pt} {\kern 1pt} {\kern 1pt} \qquad \qquad \qquad \qquad \qquad \qquad \quad~~ s = 0
\end{array} \right.
\end{align}
Suppose that the hypotheses of  sending   symbol ``1''  and  ``0''  are denoted   by  ${\cal H}_1$ and   ${\cal H}_0$, respectively.  Following  \cite{kay1993fundamentals},  the BER for the  IRS symbol assuming  maximum likelihood (ML) detection can be expressed as\footnote{The optimal estimator is a maximum a posteriori probability (MAP) detector, while 
	for equally likely symbols, the ML detector is equivalent to the  MAP detector.}
\begin{align}
{\bar P_{csr,e}} &= \frac{1}{2}\Pr \left( {{{\left| {{{\bf{v}}^H}{\rm{diag}}\left( {{\bf{h}}_r^H} \right){\bf{Gw}}} \right|}^2}\sum\limits_{l = 1}^L {{{\left| {x\left[ l \right]} \right|}^2}}+{{\bar n}_r} < \bar y_{csr,r}^{th}|{{\cal H}_1}} \right)  \notag\\
	&  + \frac{1}{2}\Pr \left( {{{\bar n}_r} \ge \bar y_{csr,r}^{th}|{{\cal H}_0}} \right),
\end{align}
where $\bar y_{csr,r}^{th} = {\left| {{{\bf{v}}^H}{\rm{diag}}\left( {{\bf{h}}_r^H} \right){\bf{Gw}}} \right|^2}\sum\limits_{l = 1}^L {{{\left| {x\left[ l \right]} \right|}^2}} /2$.
Define  probability density function (PDF) ${f_{{{\bar n}_r}}}\left( t \right) = \frac{1}{{\sqrt {2\pi } {\sigma _1}}}\exp \left( { - \frac{{{t^2}}}{{2\sigma _1^2}}} \right)$, where
 $\sigma _1^2 = {\left| {{{\bf{v}}^H}{\rm{diag}}\left( {{\bf{h}}_r^H} \right){\bf{Gw}}} \right|^2}\sum\limits_{l = 1}^L {{{\left| {x\left[ l \right]} \right|}^2}{\sigma ^2}} /2$.
Then, the BER for CSR is  obtained as 
\begin{align}
	{\bar P_{csr,e}}& = \frac{1}{{\sqrt {2\pi } {\sigma _1}}}\int_{\bar y_{csr,r}^{th}}^{ + \infty } {\exp \left( { - \frac{{{t^2}}}{{2\sigma _1^2}}} \right)} dt\notag\\
	&   = Q\left( {\frac{1}{{\sqrt 2 }}\left| {{{\bf{v}}^H}{\rm{diag}}\left( {{\bf{h}}_r^H} \right){\bf{Gw}}} \right|\sqrt {\sum\nolimits_{l = 1}^L {{{\left| {x\left[ l \right]} \right|}^2}} } /\sigma } \right),\label{csr_ber}
\end{align}
where      $Q\left( x \right) = \frac{1}{{\sqrt {2\pi } }}\int_x^\infty  {{e^{ - \frac{{{t^2}}}{2}}}} dt$. 
Since $x[l]$ in \eqref{csr_ber}  is a random variable, we are interested in the  average BER for CSR. The  primary signals are CSCG random variables and are   independent identically
distributed (i.i.d.), \eqref{csr_ber} is equivalent to the instantaneous BER for   MRC  combining of $L$ i.i.d.  Rayleigh fading paths. Hence, 
according to \cite{Proakis1995digital}, the closed-form expression for the average BER can be obtained as 
\begin{align}
{P_{csr,e}}\overset{\triangle}{=}	{\mathbb E}_{x[l]}\left\{ {{{\bar P}_{csr,e}}} \right\}={\left( {\frac{{1{\rm{ - }}\mu }}{2}} \right)^L}\sum\limits_{l = 0}^{L - 1} {\left( {\begin{array}{*{20}{c}}
		{L - 1 + l}\\
		l
		\end{array}} \right)} {\left( {\frac{{1 + \mu }}{2}} \right)^l},  \label{average_csr_ber}
\end{align}
where $\mu  = \sqrt {\frac{{{{\left| {{{\bf{v}}^H}{\rm{diag}}\left( {{\bf{h}}_r^H} \right){\bf{Gw}}} \right|}^2}}}{{{{\left| {{{\bf{v}}^H}{\rm{diag}}\left( {{\bf{h}}_r^H} \right){\bf{Gw}}} \right|}^2} + 4{\sigma ^2}}}} $ and $\left( {\begin{array}{*{20}{c}}
	n\\
	k
	\end{array}} \right) = \frac{{n\left( {n - 1} \right) \cdots \left( {n - k + 1} \right)}}{{k\left( {k - 1} \right) \cdots 1}}$. 

\subsubsection{PSR scenario}Different from the CSR scenario, for PSR, the symbol rate of  the IRS is equal to that of  the primary  transmission, i.e., $T_s=T_x$. Therefore, for PSR, the   detection  scheme  for decoding the  primary symbol  and  the IRS symbol  is  fundamentally  different from that for  CSR.
Define by $s[l]$ the IRS's $l$th  transmit symbol. The $l$th  received symbol at  the IR   is given by
\begin{align}
{y_{psr,r}}\left[ l \right] = \underbrace {{\bf{h}}_d^H{\bf{w}}x\left[ l \right]}_{{\rm{direct}}{\kern 1pt} {\kern 1pt} {\rm{link}}} + \underbrace {{\bf{h}}_r^H\left( {s\left[ l \right]{\bf{\Theta }}} \right){\bf{Gw}}x\left[ l \right]}_{{\rm{reflected}}{\kern 1pt} {\kern 1pt} {\rm{link}}} + {n_r[l]}. \label{psrcombinedsignal}
\end{align}
Similar to the CSR scenario, we first decode   the primary symbol, i.e., $x[l]$, then  subtract ${\bf{h}}_d^H{\bf{w}}x\left[ l \right]$
from the combined signal, and finally  extract the IRS symbol $s[l]$.  Since $x[l]$ and $s[l]$ have the same symbol rate for  PSR,   the IRS  treats the signal reflected from the  IRS  as interference with the average power given by ${\mathbb E}\left\{ {{{\left| {{\bf{h}}_r^H\left( {s\left[ l \right]{\bf{\Theta }}} \right){\bf{Gw}}x\left[ l \right]} \right|}^2}} \right\} = \rho {\left| {{{\bf{v}}^H}{\rm{diag}}\left( {{\bf{h}}_r^H} \right){\bf{Gw}}} \right|^2}$ when decoding the primary signal $x[l]$. 
Therefore, the   achievable rate for decoding the primary signal is given by 
\begin{align}
{R_{psr,x}} = {\log _2}\left( {1 + \frac{{{{\left| {{\bf{h}}_d^H{\bf{w}}} \right|}^2}}}{{\rho {{\left| {{{\bf{v}}^H}{\rm{diag}}\left( {{\bf{h}}_r^H} \right){\bf{Gw}}} \right|}^2}{\rm{ + }}{\sigma ^2}}}} \right),
\end{align}
and then after removing ${\bf{h}}_d^H{\bf{w}}x\left[ l \right]$ from  \eqref{psrcombinedsignal}, we have
\begin{align}
{\bar y_{psr,r}}\left[ l \right] = {\bf{h}}_r^H{\bf{\Theta Gw}}s\left[ l \right]x\left[ l \right] + {n_r[l]}.\label{psrsubstractsignal}
\end{align}
Similar to the case of CSR, by setting $L=1$ in \eqref{average_csr_ber}, the average  BER for PSR  can be expressed as 
\begin{align}
{P_{psr,e}} = \frac{1}{2}{\rm{ - }}\frac{1}{2}\sqrt {\frac{{{{\left| {{{\bf{v}}^H}{\rm{diag}}\left( {{\bf{h}}_r^H} \right){\bf{Gw}}} \right|}^2}}}{{{{\left| {{{\bf{v}}^H}{\rm{diag}}\left( {{\bf{h}}_r^H} \right){\bf{Gw}}} \right|}^2} + 4{\sigma ^2}}}}.\label{average_psr_ber}
\end{align}


\subsection{Problem Formulation}
\subsubsection{CSR scenario}
Our goal is to minimize the  BER of the  IRS symbols by jointly optimizing the IRS phase shifts  and the BS transmit beamformer while guaranteeing a minimum rate required for the primary network  subject to the BS transmit power budget and the unit-modulus phase-shift constraints.  Mathematically, the problem can be formulated as  follows
\begin{align}
&\left( {{\rm{P}}1} \right)\mathop {\min }\limits_{{\bf{w}},{\bf{v}}} {\left( {\frac{{1{\rm{ - }}\mu }}{2}} \right)^L}\sum\limits_{l = 0}^{L - 1} {\left( {\begin{array}{*{20}{c}}
		{L - 1 + l}\\
		l
		\end{array}} \right)} {\left( {\frac{{1 + \mu }}{2}} \right)^l}\notag\\
&{\rm{s}}{\rm{.t}}{\rm{.}}~ \left( {1 - \rho } \right){\log _2}\left( {1 + \frac{{{{\left| {{\bf{h}}_d^H{\bf{w}}} \right|}^2}}}{{{\sigma ^2}}}} \right) + \rho {\log _2}\left( {1 + \frac{{{{\left| {{\bf{h}}_d^H{\bf{w}} + {{\bf{v}}^H}{\rm{diag}}\left( {{\bf{h}}_r^H} \right){\bf{Gw}}} \right|}^2}}}{{{\sigma ^2}}}} \right) \ge {R_{csr,{\rm th}}},\label{csrP1const1}\\
&\qquad \left\| {\bf{w}} \right\|_2^2 \le {P_{\max }},\label{csrP1const2}\\
&\qquad \left| {{{\bf{v}}_m}} \right| = 1,\forall m,\label{csrP1const3}
\end{align}
where  ${{{\bf{v}}_m}}$ denotes the $m$th element of  ${\bf{v}}$, ${R_{csr,{\rm{th}}}}$  is the minimum rate required  by  the primary network  for   CSR, and ${P_{\max }}$ is   the BS's maximum transmit power.
Problem $\left( {{\rm{P}}1} \right)$ is non-convex and difficult to  solve due to   the highly coupled optimization variables in the objective function as well as  constraints \eqref{csrP1const1}  and \eqref{csrP1const3}. There is no standard
method for solving   non-convex optimization problems optimally. As such, we propose  
a novel penalty-based algorithm  to solve $\left( {{\rm{P}}1} \right)$  to obtain  a high-quality suboptimal solution in  Section III.
\subsubsection{PSR scenario}
Similarly, for PSR, we aim to jointly optimize the IRS
phase shifts and the BS transmit beamformer to minimize the BER. Accordingly, the  problem can be formulated as
\begin{align}
&\left( {{\rm{P}}2} \right)\mathop {\min }\limits_{{\bf{w}},{\bf{v}}} \frac{1}{2}{\rm{ - }}\frac{1}{2}\sqrt {\frac{{{{\left| {{{\bf{v}}^H}{\rm{diag}}\left( {{\bf{h}}_r^H} \right){\bf{Gw}}} \right|}^2}}}{{{{\left| {{{\bf{v}}^H}{\rm{diag}}\left( {{\bf{h}}_r^H} \right){\bf{Gw}}} \right|}^2} + 4{\sigma ^2}}}} \notag\\
&{\rm{s}}{\rm{.t}}{\rm{.}}{\kern 1pt} {\kern 1pt} {\kern 1pt} {\kern 1pt} {\kern 1pt} {\kern 1pt} {\log _2}\left( {1 + \frac{{{{\left| {{\bf{h}}_d^H{\bf{w}}} \right|}^2}}}{{\rho {{\left| {{{\bf{v}}^H}{\rm{diag}}\left( {{\bf{h}}_r^H} \right){\bf{Gw}}} \right|}^2} + {\sigma ^2}}}} \right) \ge {R_{psr,{\rm{th}}}},\label{psrP2const1}\\
&\qquad \eqref{csrP1const2}, \eqref{csrP1const3},
\end{align}
where ${R_{psr,{\rm{th}}}}$ represents the minimum rate required  for the  primary network for  PSR. Problem $\left( {{\rm{P}}2} \right)$ is also challenging  to solve for the following three  reasons. First, 
the objective function  of  $\left( {{\rm{P}}2} \right)$  is  rather complicated, as it is non-convex due to the involvement of  coupled optimization variables $\bf w$ and $\bf v$. Second, the optimization variables are intricately coupled in constraint \eqref{psrP2const1}. Third, constraint \eqref{csrP1const3} is a   unit-modulus constraint. Nevertheless, we propose   an efficient  bisection search based algorithm to solve problem  $\left( {{\rm{P}}2} \right)$ in Section IV.

\section{Penalty-based Algorithm for CSR Optimization  Problem}
In this section,   we study the CSR scenario to minimize the BER of the IRS symbols and  propose a novel penalty-based algorithm to solve $\left( {{\rm{P}}1} \right)$, which involves a two-layer iteration, i.e.,  an inner layer iteration and an outer layer iteration. Specifically, the inner  layer solves the penalized optimization problem, while the  outer layer updates the penalty coefficient. We then alternately  optimize the two layer iterations until  convergence is achieved. Before proceeding to solving the problem, 
we observe that ${P_{csr,e}}$ given by \eqref{average_csr_ber} is monotonically decreasing in ${{{\left| {{{\bf{v}}^H}{\rm{diag}}\left( {{\bf{h}}_r^H} \right){\bf{Gw}}} \right|^2}}}$, which implies that    minimizing  the BER $P_{csr,e}$ is equivalent to maximizing the SNR, i.e., $\frac{{{{\left| {{{\bf{v}}^H}{\rm{diag}}\left( {{\bf{h}}_r^H} \right){\bf{Gw}}} \right|}^2}}}{{{\sigma ^2}}}$. Thus, in the following, we   adopt  the SNR as the objective function to facilitate  algorithm design.
\subsection{Problem Reformulation}
 We first introduce  new auxiliary variables ${{\mu _1}}$ and ${{\mu _2}}$ satisfying  ${\mu _1}{\rm{ = }}\frac{{{{\bf{v}}^H}{\rm{diag}}\left( {{\bf{h}}_r^H} \right){\bf{Gw}}}}{\sigma }$ and ${\mu _2} = \frac{{{\bf{h}}_d^H{\bf{w}}}}{\sigma }$. Then, problem $\left( {{\rm{P}}1} \right)$ is equivalent to
\begin{align}
\left( {{\rm{\bar P}}1} \right)& \mathop {\max }\limits_{{\bf{w}},{\bf{v}},{\mu _1},{\mu _2}} {\left| {{\mu _1}} \right|^2}\notag\\
&{\rm{s}}{\rm{.t}}{\rm{.}}{\kern 1pt} {\kern 1pt} {\kern 1pt} {\kern 1pt} {\kern 1pt} {\kern 1pt} \left( {1 - \rho } \right){\log _2}\left( {1 + {{\left| {{\mu _2}} \right|}^2}} \right) + \rho {\log _2}\left( {1 + {{\left| {{\mu _1} + {\mu _2}} \right|}^2}} \right) \ge {R_{csr,{\rm{th}}}},\label{csrbarP1const1}\\
&\qquad{\mu _1}{\rm{ = }}\frac{{{{\bf{v}}^H}{\rm{diag}}\left( {{\bf{h}}_r^H} \right){\bf{Gw}}}}{\sigma },\label{csrbarP1const2}\\
&\qquad{\mu _2} = \frac{{{\bf{h}}_d^H{\bf{w}}}}{\sigma },\label{csrbarP1const3}\\
&\qquad \eqref{csrP1const2},\eqref{csrP1const3}.
\end{align}
We then  use \eqref{csrbarP1const2} and \eqref{csrbarP1const3} as   penalty terms that are added to the objective function of $\left( {{\rm{\bar P}}1} \right)$, yielding the following optimization problem
\begin{align}
\left( {{\rm{\bar P}}1{\rm{ - }}1} \right)& \mathop {\max }\limits_{{\bf{w}},{\bf{v}},{\mu _1},{\mu _2}} {\left| {{\mu _1}} \right|^2} - \frac{1}{{2\eta }}\left( {{{\left| {{\mu _1} - \frac{{{{\bf{v}}^H}{\rm{diag}}\left( {{\bf{h}}_r^H} \right){\bf{Gw}}}}{\sigma }} \right|}^2} + {{\left| {{\mu _2} - \frac{{{\bf{h}}_d^H{\bf{w}}}}{\sigma }} \right|}^2}} \right)\\
&{\rm{s}}{\rm{.t}}{\rm{.}}{\kern 1pt} {\kern 1pt} {\kern 1pt} {\kern 1pt} {\kern 1pt} {\kern 1pt}  \eqref{csrP1const2},\eqref{csrP1const3},\eqref{csrbarP1const1},
\end{align}
where $\eta $ ($\eta  > 0$) is a penalty coefficient that penalizes the violation of equality constraints  \eqref{csrbarP1const2} and \eqref{csrbarP1const3}. By gradually decreasing the value of
 $\eta $ in the outer layer until  it approaches  zero, it follows that $\frac{1}{{2\eta }} \to \infty $.  As such, the penalty terms will be forced to   zero eventually,   i.e., $\left| {{\mu _1} - \frac{{{{\bf{v}}^H}{\rm{diag}}\left( {{\bf{h}}_r^H} \right){\bf{Gw}}}}{\sigma }} \right| = 0$ and $\left| {{\mu _2} - \frac{{{\bf{h}}_d^H{\bf{w}}}}{\sigma }} \right| = 0$, which indicates that  the newly added equality constraints \eqref{csrbarP1const2} and \eqref{csrbarP1const3} will be  satisfied after convergence. However,   for a given $\eta $, $\left( {{\rm{\bar P}}1{\rm{ - }}1} \right)$   is still a non-convex optimization problem due to the coupled optimization variables in both the objective function and  constraint  \eqref{csrbarP1const1}, and the unit-modulus constraint in \eqref{csrP1const3}. To address this difficulty, we first  divide the  variables into three blocks, namely, 1) auxiliary variables $\{\mu _1,\mu _2\}$, 2) BS transmit beamformer $\bf w$, and 3) phase shift vector $\bf v$, and   alternately optimize each block with the other two blocks of variables fixed  until   convergence is achieved.
\subsection{Inner Layer Iteration}
\textit{1) Optimizing auxiliary variables $\{\mu _1,\mu _2\}$ for the given BS transmit beamformer  $\bf w$ and phase shift vector $\bf v$.} This subproblem is formulated as 
\begin{align}
\left( {{\rm{\bar P}}1{\rm{ - 2}}} \right)& \mathop {\max }\limits_{{\mu _1},{\mu _2}} ~{\left| {{\mu _1}} \right|^2} - \frac{1}{{2\eta }}\left( {{{\left| {{\mu _1} - \frac{{{{\bf{v}}^H}{\rm{diag}}\left( {{\bf{h}}_r^H} \right){\bf{Gw}}}}{\sigma }} \right|}^2} + {{\left| {{\mu _2} - \frac{{{\bf{h}}_d^H{\bf{w}}}}{\sigma }} \right|}^2}} \right)\notag\\
&{\rm{s}}{\rm{.t}}{\rm{.}}{\kern 1pt} {\kern 1pt} {\kern 1pt} {\kern 1pt} {\kern 1pt} {\kern 1pt}\eqref{csrbarP1const1}.
\end{align}
Note that $\left( {{\rm{\bar P}}1{\rm{ - 2}}} \right)$ is neither concave nor quasi-concave due to the non-convex constraint \eqref{csrbarP1const1}. In addition,  we must have $1{\rm{ - }}\frac{1}{{2\eta }} \le 0$ in the objective function, i.e., $\eta  \le \frac{1}{2}$,  since otherwise we can  increase $\mu_1$ to obtain an infinite  value of the objective function. As such, the objective function of $\left( {{\rm{\bar P}}1{\rm{ - 2}}} \right)$ is jointly  convex w.r.t. $\mu_1$ and $\mu_2$.
In the following, we propose to leverage the successive convex approximation (SCA) technique to solve $\left( {{\rm{\bar P}}1{\rm{ - 2}}} \right)$. Recall that any convex function is globally lower-bounded by its first-order Taylor expansion at any feasible point. Therefore, for any given points $\mu^r_1$ and $\mu^r_2$ at the $r$th iteration,  we have
\begin{align}
{\left| {{\mu _2}} \right|^2} \ge { - {{\left| {\mu _2^r} \right|}^2} + 2{\rm{Re}}\left\{ {\mu _2^H\mu _2^r} \right\}},\label{csr_p1_6_const3_new}
\end{align}
\begin{align}
{\left| {{\mu _1}{\rm{ + }}{\mu _2}} \right|^2}\ge { - {{\left| {\mu _1^r + \mu _2^r} \right|}^2} + 2{\rm{Re}}\left\{ {{{\left( {{\mu _1} + {\mu _2}} \right)}^H}\left( {\mu _1^r + \mu _2^r} \right)} \right\}}.\label{csr_p1_6_const4_new}
\end{align}
As a result, for any given  points $\mu^r_1$ and $\mu^r_2$, we obtain the following optimization problem 
\begin{align}
\left( {{\rm{\bar P}}1{\rm{ - 3}}} \right)& \mathop {\max }\limits_{{\mu _1},{\mu _2}}~{\left| {{\mu _1}} \right|^2} - \frac{1}{{2\eta }}\left( {{{\left| {{\mu _1} - \frac{{{{\bf{v}}^H}{\rm{diag}}\left( {{\bf{h}}_r^H} \right){\bf{Gw}}}}{\sigma }} \right|}^2} + {{\left| {{\mu _2} - \frac{{{\bf{h}}_d^H{\bf{w}}}}{\sigma }} \right|}^2}} \right)\notag\\
&{\rm{s}}{\rm{.t}}{\rm{.}}{\kern 1pt} {\kern 1pt} {\kern 1pt} {\kern 1pt} {\kern 1pt} {\kern 1pt} \left( {1 - \rho } \right){\log _2}\left( {1 - {{\left| {\mu _2^r} \right|}^2} + 2{\rm{Re}}\left\{ {\mu _2^H\mu _2^r} \right\}} \right) + \notag\\
&\qquad \rho {\log _2}\left( {1 - {{\left| {\mu _1^r + \mu _2^r} \right|}^2} + 2{\rm{Re}}\left\{ {{{\left( {{\mu _1} + {\mu _2}} \right)}^H}\left( {\mu _1^r + \mu _2^r} \right)} \right\}} \right) \ge {R_{csr,{\rm{th}}}}. \label{csr_p1_7_const1}
\end{align}
It can be readily verified that the objective function of $\left( {{\rm{\bar P}}1{\rm{ - 3}}} \right)$ and  new constraint \eqref{csr_p1_7_const1} are convex. Thus, $\left( {{\rm{\bar P}}1{\rm{ - 3}}} \right)$ can be efficiently solved by using  standard convex optimization techniques \cite{boyd2004convex}. It is worth pointing out that the obtained objective value of 
$\left( {{\rm{\bar P}}1{\rm{ - 3}}} \right)$ serves  as a lower-bound for $\left( {{\rm{\bar P}}1{\rm{ - 2}}} \right)$  due to the Taylor expansion approximation  in   \eqref{csr_p1_6_const3_new} and \eqref{csr_p1_6_const4_new}.

\hspace*{\parindent}\textit{2) Optimizing BS transmit beamformer  $\bf w$ for the given  phase shift  $\bf v$ and auxiliary variables $\{\mu _1,\mu _2\}$.}  This subproblem can be expressed as
\begin{align}
\left( {{\rm{\bar P}}1{\rm{ - 4}}} \right)& \mathop {\min }\limits_{\bf{w}} {\left| {{\mu _1} - \frac{{{{\bf{v}}^H}{\rm{diag}}\left( {{\bf{h}}_r^H} \right){\bf{Gw}}}}{\sigma }} \right|^2} + {\left| {{\mu _2} - \frac{{{\bf{h}}_d^H{\bf{w}}}}{\sigma }} \right|^2}\notag\\
&{\rm{s}}{\rm{.t}}{\rm{.}}{\kern 1pt} {\kern 1pt} {\kern 1pt} {\kern 1pt} {\kern 1pt} {\kern 1pt}  \eqref{csrP1const2}.
\end{align}
It can be readily observed  that $\left( {{\rm{\bar P}}1{\rm{ - 4}}} \right)$ is a convex  quadratically constrained quadratic program (QCQP), which can be solved by the interior point method \cite{boyd2004convex}. However, the complexity of   solving  $\left( {{\rm{\bar P}}1{\rm{ - 4}}} \right)$  by the interior point method is ${\cal O}{\left( N^{3.5} \right)}$, which is rather  high especially when the number of antennas $N$ is  large. To reduce the computational complexity, we obtain  a  semi-closed-form yet optimal  solution for  the BS transmit beamformer  $\bf w$ by using the Lagrange duality  method \cite{boyd2004convex}. Specifically, by introducing dual variable $\lambda $ ($\lambda  \ge 0$) associated with constraint \eqref{csrP1const2}, the  Lagrangian function of $\left( {{\rm{\bar P}}1{\rm{ - 4}}} \right)$ is given by 
\begin{align}
{\cal L}\left( {{\bf{w}},\lambda } \right) = {\left| {{\mu _1} - \frac{{{{\bf{v}}^H}{\rm{diag}}\left( {{\bf{h}}_r^H} \right){\bf{Gw}}}}{\sigma }} \right|^2} + {\left| {{\mu _2} - \frac{{{\bf{h}}_d^H{\bf{w}}}}{\sigma }} \right|^2} + \lambda \left( {\left\| {\bf{w}} \right\|_2^2 - {P_{\max }}} \right).
\end{align}
By taking  the first-order derivative of ${\cal L}\left( {{\bf{w}},\lambda } \right)$ w.r.t. $\bf w$  and setting it to zero, we  obtain the optimal solution as 
\begin{align}
{\bf{w}}\left( \lambda  \right){\rm{ = }}{\left( {\frac{{{{\bf{G}}^H}{\rm{diag}}\left( {{{\bf{h}}_r}} \right){\bf{v}}{{\bf{v}}^H}{\rm{diag}}\left( {{\bf{h}}_r^H} \right){\bf{G}}{\rm{ + }}{{\bf{h}}_d}{\bf{h}}_d^H}}{{{\sigma ^2}}}{\rm{ + }}\lambda {{\bf{I}}_N}} \right)^\dag }\left( {\frac{{{\mu _1}{{\bf{G}}^H}{\rm{diag}}\left( {{{\bf{h}}_r}} \right){\bf{v}}{\rm{ + }}{\mu _2}{{\bf{h}}_d}}}{\sigma }} \right). \label{csr_beamforming}
\end{align}
Recall that for the optimal solution ${{\bf{w}}^{{\rm{opt}}}}\left( {{\lambda ^{{\rm{opt}}}}} \right)$ and ${\lambda ^{{\rm{opt}}}}$, the following complementary slackness condition must be satisfied\cite{boyd2004convex}
\begin{align}
{\lambda ^{{\rm{opt}}}}\left( {{{\left\| {{{\bf{w}}^{{\rm{opt}}}}\left( {{\lambda ^{{\rm{opt}}}}} \right)} \right\|}^2} - {P_{\max }}} \right) = 0.
\end{align}
We first check whether ${\lambda ^{{\rm{opt}}}} = 0$ is the optimal solution or not. If
\begin{align}
{\left\| {{{\bf{w}}^{{\rm{opt}}}}\left( 0 \right)} \right\|^2} - {P_{\max }} < 0,
\end{align}
which indicates that the optimal dual variable $\lambda$ equals   $0$, the optimal BS beamformer   is given by ${{\bf{w}}^{{\rm{opt}}}}\left( 0 \right){\rm{ = }}{\left( {{{{{\bf{G}}^H}{\rm{diag}}\left( {{{\bf{h}}_r}} \right){\bf{v}}{{\bf{v}}^H}{\rm{diag}}\left( {{\bf{h}}_r^H} \right){\bf{G}}{\rm{ + }}{{\bf{h}}_d}{\bf{h}}_d^H} \mathord{\left/
				{\vphantom {{{{\bf{G}}^H}{\rm{diag}}\left( {{{\bf{h}}_r}} \right){\bf{v}}{{\bf{v}}^H}{\rm{diag}}\left( {{\bf{h}}_r^H} \right){\bf{G}}{\rm{ + }}{{\bf{h}}_d}{\bf{h}}_d^H} {{\sigma ^2}}}} \right.
				\kern-\nulldelimiterspace} {{\sigma ^2}}}} \right)^ {\dag} }\left( {\frac{{{\mu _1}{{\bf{G}}^H}{\rm{diag}}\left( {{{\bf{h}}_r}} \right){\bf{v}}{\rm{ + }}{\mu _2}{{\bf{h}}_d}}}{\sigma }} \right)$, otherwise, the optimal $\lambda$ is a positive value, which can  be calculated  as follows.
						
\noindent	 Defining ${\bf{S}} = {{\bf{G}}^H}{\rm{diag}}\left( {{{\bf{h}}_r}} \right){\bf{v}}{{\bf{v}}^H}{\rm{diag}}\left( {{\bf{h}}_r^H} \right){\bf{G}}{\rm{ + }}{{\bf{h}}_d}{\bf{h}}_d^H/{\sigma ^2}$ and ${\bf{z}} = {\mu _1}{{\bf{G}}^H}{\rm{diag}}\left( {{{\bf{h}}_r}} \right){\bf{v}}{\rm{ + }}{\mu _2}{{\bf{h}}_d}/\sigma $,  
we have 
\begin{align}
\left\| {{\bf{w}}\left( \lambda  \right)} \right\|_2^2{\rm{ = tr}}\left( {{{\left( {{\bf{S}}{\rm{ + }}\lambda {{\bf{I}}_N}} \right)}^{ - 2}}{\bf{z}}{{\bf{z}}^H}} \right).\label{csr_beamforing_lamba}
\end{align}
It can be readily shown  that  ${\bf{S}}$ is a positive semi-definite matrix.  We thus have ${\bf{S}}{\rm{ = }}{\bf{U\Sigma }}{{\bf{U}}^H}$ by  performing eigenvalue decomposition. Substituting ${\bf{S}}{\rm{ = }}{\bf{U\Sigma }}{{\bf{U}}^H}$ into  \eqref{csr_beamforing_lamba}, we arrive at 
\begin{align}
\left\| {{\bf{w}}\left( \lambda  \right)} \right\|_2^2&{\rm{ = tr}}\left( {{{\left( {{\bf{\Sigma }}{\rm{ + }}\lambda {{\bf{I}}_N}} \right)}^{ - 2}}{{\bf{U}}^H}{\bf{z}}{{\bf{z}}^H}{\bf{U}}} \right)\notag\\
&{\rm{ = }}\sum\limits_{i = 1}^{{s_r}} {\frac{{{{\left( {{{\bf{U}}^H}{\bf{z}}{{\bf{z}}^H}{\bf{U}}} \right)}_{i,i}}}}{{{{\left( {{{\bf{\Sigma }}_{i,i}} + \lambda } \right)}^2}}}}  + \sum\limits_{i = {s_r} + 1}^N {\frac{{{{\left( {{{\bf{U}}^H}{\bf{z}}{{\bf{z}}^H}{\bf{U}}} \right)}_{i,i}}}}{{{\lambda ^2}}}}, 
\end{align}
where $s_r$ denotes the number of  non-zero eigenvalues of ${\bf{S}}$. Note that since each  main diagonal element of $ {{{\bf{U}}^H}{\bf{z}}{{\bf{z}}^H}{\bf{U}}}$ is  non-negative, $\left\| {{\bf{w}}\left( \lambda  \right)} \right\|_2^2$ is   monotonically decreasing w.r.t. dual variable $\lambda $. Therefore, the optimal ${{\lambda ^{{\rm{opt}}}}}$ can be obtained by using the bisection
 search method to find the solution  that satisfies  the following equation
\begin{align}
\sum\limits_{i = 1}^{{s_r}} {\frac{{{{\left( {{{\bf{U}}^H}{\bf{z}}{{\bf{z}}^H}{\bf{U}}} \right)}_{i,i}}}}{{{{\left( {{{\bf{\Sigma }}_{i,i}} + {\lambda ^{{\rm{opt}}}}} \right)}^2}}}}  + \sum\limits_{i = {s_r} + 1}^N {\frac{{{{\left( {{{\bf{U}}^H}{\bf{z}}{{\bf{z}}^H}{\bf{U}}} \right)}_{i,i}}}}{{{{\left( {{\lambda ^{{\rm{opt}}}}} \right)}^2}}}}  =  {P_{\max }}.
\end{align}
Then, substituting the optimal ${{\lambda ^{{\rm{opt}}}}}$ into \eqref{csr_beamforming}, we   obtain the optimal BS beamformer vector ${{\bf{w}}^{{\rm{opt}}}}\left( {{\lambda ^{{\rm{opt}}}}} \right)$. Note that for the bisection search,  the low bound of $\lambda $, denoted by $\lambda^{\rm lb} $, can be set as a sufficiently small non-negative  value, while 
 the upper bound of $\lambda $, denoted by $\lambda^{\rm up} $, can be calculated as  ${\lambda ^{{\rm{up}}}} = \sqrt {\sum\limits_{i = 1}^N {{{\left( {{{\bf{U}}^H}{\bf{z}}{{\bf{z}}^H}{\bf{U}}} \right)}_{i,i}}} /{P_{\max }}} $.
The  detailed procedure  for solving  $\left( {{\rm{\bar P}}1{\rm{ - 4}}} \right)$ is  summarized in Algorithm~\ref{alg1}. Note that  the complexity of Algorithm~\ref{alg1} is ${\cal O}\left( {\log_2 \left( {\frac{{{\lambda ^{{\rm{up}}}} - {\lambda ^{{\rm{lb}}}}}}{\varepsilon }} \right)}N^3 \right)$, which is much lower than that of  the interior point method.
\begin{algorithm}[!t]
	\caption{Proposed Lagrange  duality  method  for solving problem $\left( {{\rm{\bar P}}1{\rm{ - 4}}} \right)$.}
	\label{alg1}
	\begin{algorithmic}[1]
		\STATE  \textbf{Initialize} ${\lambda ^{{\rm{lb}}}}$, ${\lambda ^{{\rm{up}}}}$, and  $\varepsilon$.
		\STATE  \textbf{If} $\left\| {{\bf{w}}\left( 0 \right)} \right\|_2^2 \le {P_{\max }}$, the optimal BS beamforming vector is given by ${{{\bf{w}}^{{\rm{opt}}}}\left( 0 \right)}$, and then  terminate the algorithm;
		otherwise,  go to step $3$.
			\STATE \textbf{Repeat}
		\STATE \quad Compute $\lambda  = \frac{{{\lambda ^{{\rm{lb}}}} + {\lambda ^{{\rm{up}}}}}}{2}$. 

		\STATE \quad If $\left\| {{\bf{w}}\left( \lambda  \right)} \right\|_2^2 \le {P_{\max }}$, let ${\lambda ^{{\rm{up}}}} = \lambda $, otherwise, let ${\lambda ^{{\rm{lb}}}} = \lambda $.
		\STATE \textbf{Until} $\left| {{\lambda ^{{\rm{up}}}} - {\lambda ^{{\rm{lb}}}}} \right| \le \varepsilon $.
		\STATE {\bf Output:}  Optimal BS beamformer ${{\bf{w}}^{{\rm{opt}}}}\left( {{\lambda ^{{\rm{lb}}}}} \right)$.
	\end{algorithmic}
\end{algorithm}

\hspace*{\parindent}\textit{3) Optimizing phase shift vector $\bf v$ for   given BS transmit beamformer  $\bf w$ and auxiliary variables $\{\mu _1,\mu _2\}$.} This subproblem is given by 
\begin{align}
\left( {{\rm{\bar P}}1{\rm{ - 5}}} \right)& \mathop {\min }\limits_{\bf{v}} {\left| {{\mu _1} - \frac{{{{\bf{v}}^H}{\rm{diag}}\left( {{\bf{h}}_r^H} \right){\bf{Gw}}}}{\sigma }} \right|^2}\notag\\
&{\rm{s}}{\rm{.t}}{\rm{.}}{\kern 1pt} {\kern 1pt} {\kern 1pt} {\kern 1pt} {\kern 1pt} {\kern 1pt}\eqref{csrP1const3}.
\end{align}
Although the objective function of $\left( {{\rm{\bar P}}1{\rm{ - 5}}} \right)$ is a quadratic function, the unit-modulus constraints in \eqref{csrP1const3} are still  non-convex. In the following, we obtain     a locally optimal closed-form solution  for $\bf v$ by leveraging the Majorization-Minimization (MM) method\cite{song2016Sequence,sun2016majorization}. 
The key idea of using the MM method lies in   constructing  a convex surrogate function  that is an upper bound of the objective function of $\left( {{\rm{\bar P}}1{\rm{ - 5}}} \right)$. Define $f\left( {\bf{v}} \right)$ as the objective function   and ${{{\bf{v}}^r}}$ as the initial point for ${\bf{v}}$ at the $r$th iteration, the  surrogate function, denoted by $\hat f\left( {{\bf{v}}{\rm{|}}{{\bf{v}}^r}} \right)$, should  satisfy with the following three conditions: 1) $\hat f\left( {{\bf{v}}{\rm{|}}{{\bf{v}}^r}} \right) \ge f\left( {\bf{v}} \right)$; 2) $\hat f\left( {{{\bf{v}}^r}{\rm{|}}{{\bf{v}}^r}} \right) = f\left( {{{\bf{v}}^r}} \right)$;
3) ${\nabla _{{{\bf{v}}^r}}}\hat f\left( {{{\bf{v}}^r}{\rm{|}}{{\bf{v}}^r}} \right) = {\nabla _{{{\bf{v}}^r}}}f\left( {{{\bf{v}}^r}} \right)$, where 1) means that  $\hat f\left( {{\bf{v}}{\rm{|}}{{\bf{v}}^r}} \right)$   serves   an upper bound   function of $f\left( {\bf{v}} \right)$, 2) implies that  $\hat f\left( {{\bf{v}}{\rm{|}}{{\bf{v}}^r}} \right)$ and $f\left( {\bf{v}} \right)$ have the same function value  at  point ${{{\bf{v}}^r}}$, and 3) indicates $\hat f\left( {{\bf{v}}{\rm{|}}{{\bf{v}}^r}} \right)$ and $f\left( {\bf{v}} \right)$ have the same gradient at  point ${{{\bf{v}}^r}}$. As a result, we have the following lemma:

\textbf{\emph{Lemma 1:}} Based on \cite{song2016Sequence}, at the initial point ${{{\bf{v}}^r}}$, a surrogate function $\hat f\left( {{\bf{v}}{\rm{|}}{{\bf{v}}^r}} \right)$ for quadratic function $f({\bf v})={{\bf{v}}^H}{\bf{Av}}$  is given by
\begin{align}
\hat f\left( {{\bf{v}}|{{\bf{v}}^r}} \right) = {\lambda _{\max }}{{\bf{v}}^H}{\bf{v}} - 2{\mathop{\rm Re}\nolimits} \left\{ {{{\bf{v}}^H}\left( {{\lambda _{\max }}{{\bf{I}}_M} - {\bf{A}}} \right){{\bf{v}}^r}} \right\} + {{\bf{v}}^{r,H}}\left( {{\lambda _{\max }}{{\bf{I}}_M} - {\bf{A}}} \right){{\bf{v}}^r}, \label{csr_f_surrgogate}
\end{align}
where ${\bf{A}} = \frac{{{\rm{diag}}\left( {{\bf{h}}_r^H} \right){\bf{Gw}}{{\bf{w}}^H}{{\bf{G}}^H}{\rm{diag}}\left( {{{\bf{h}}_r}} \right)}}{{{\sigma ^2}}}$, and ${\lambda _{\max }}$ represents  the maximum  eigenvalue of ${\bf{A}}$. 

Substituting  $\hat f\left( {{\bf{v}}|{{\bf{v}}^r}} \right)$  for   $f\left( {\bf{v}} \right)$  and plugging it  into the objective function of $\left( {{\rm{\bar P}}1{\rm{ - 5}}} \right)$ as well as  ignoring  the irrelevant constants w.r.t. $\bf v$,  the phase shift $\bf v$ can be obtained by solving the following problem 
\begin{align}
\left( {{\rm{\bar P}}1{\rm{ - 6}}} \right)& \mathop {\max }\limits_{\bf{v}} ~{\mathop{\rm Re}\nolimits} \left\{ {{{\bf{v}}^H}{{\bf{q}}^r}} \right\}\notag\\
&{\rm{s}}{\rm{.t}}{\rm{.}}{\kern 1pt} {\kern 1pt} {\kern 1pt} {\kern 1pt} {\kern 1pt} {\kern 1pt}\eqref{csrP1const3},
\end{align}
where ${{\bf{q}}^r} = \left( {{\lambda _{\max }}{{\bf{I}}_M} - {\bf{A}}} \right){{\bf{v}}^r} + \frac{{{\rm{diag}}\left( {{\bf{h}}_r^H} \right){\bf{Gw}}\mu _1^H}}{\sigma }$.
Obviously, the optimal phase shift vector for $\left( {{\rm{\bar P}}1{\rm{ - 6}}} \right)$ is given by 
${{\bf{v}}^{{\rm{opt}}}} = \exp \left( {j\arg \left( {{{\bf{q}}^r}} \right)} \right)$. Note that the   obtained optimal  solution  ${{\bf{v}}^{{\rm{opt}}}}$ for  $\left( {{\rm{\bar P}}1{\rm{ - 6}}} \right) $    is guaranteed to be  a locally optimal solution for the original problem $\left( {{\rm{\bar P}}1{\rm{ - 5}}} \right) $ \cite{song2016Sequence,sun2016majorization}.

\subsection{Outer Layer Iteration}
In the outer layer, we gradually decrease the value of  penalty coefficient $\eta^{r}$ in the $r$th iteration  by updating it  as follows
\begin{align}
\eta^r=c\eta^{r-1}, \label{penaltcoefficentupdate}
\end{align}
where $c$ $(0<c<1)$ is a  scaling factor. Here,   a larger value of $c$ can achieve
better performance but at the cost of more iterations required in the outer layer.
\subsection{Overall Algorithm}
 The constraint violation  of the proposed  penalty-based algorithm  is qualified by 
\begin{align}
\xi {\rm{ = }}\max \left\{ {\left| {{\mu _1} - \frac{{{{\bf{v}}^H}{\rm{diag}}\left( {{\bf{h}}_r^H} \right){\bf{Gw}}}}{\sigma }} \right|,\left| {{\mu _2} - \frac{{{\bf{h}}_d^H{\bf{w}}}}{\sigma }} \right|} \right\}. \label{constraintviolation}
\end{align}
The proposed  penalty-based algorithm is summarized in Algorithm~\ref{alg2}.
\begin{algorithm}[!t]
	\caption{Proposed penalty-based algorithm for solving problem $\left( {{\rm{\bar P}}1-1} \right)$.}
	\label{alg2}
	\begin{algorithmic}[1]
		\STATE  \textbf{Initialize}  phase shift vector ${\bf v}^{r_1}$, auxiliary variables $\{\mu_1^{r_1}, \mu_2^{r_1}\}$, penalty coefficient $\eta^{r_2}$,  scaling factor $c$,  predefined thresholds $\varepsilon_1$ and  $\varepsilon_2$,   inner layer iteration index  $r_1=0$,   outer layer iteration index  $r_2=0$.
		\STATE  \textbf{Repeat: outer layer}
		\STATE \quad \textbf{Repeat: inner layer }
				\STATE  \qquad Update auxiliary variables, denoted by $\{\mu_1^{r_1+1}, \mu_2^{r+1}\}$, by solving problem $\left( {{\rm{\bar P}}1{\rm{-3}}} \right)$.
		\STATE  \qquad Update BS transmit  beamformer, denoted by  ${\bf w}^{r_1+1}$, by solving problem $\left( {{\rm{\bar P}}1{\rm{-4}}} \right)$.
		\STATE  \qquad Update phase shift vector, denoted by ${\bf v}^{r_1+1}$, by solving problem $\left( {{\rm{\bar P}}1{\rm{-6}}} \right)$.
		\STATE  \qquad Set $r_1= r_1+1$.
      	\STATE \quad \textbf{Until} the fractional increase of  the objective value of $\left( {{\rm{\bar P}}1{\rm{ - 1}}} \right)$ is below  $\varepsilon_1$.
		\STATE  \quad Update the penalty coefficient, denoted by $\eta^{r_2+1}$, based on \eqref{penaltcoefficentupdate}.
		\STATE  \quad Set $r_2= r_2+1$ and $r_1= 0$.
		\STATE \textbf{Until}  constraint violation indicator $\xi $ in \eqref{constraintviolation} is smaller than   $\varepsilon_2$.
	\end{algorithmic}
\end{algorithm}

\textbf{\emph{Lemma 2:}} The obtained solution $\{{\bf w}, {\bf v}, \mu_1,\mu_2\}$ converges to a point fulfilling the Karush–Kuhn –Tucker (KKT) optimality conditions of   original problem $\left( {{\rm{\bar P}}1} \right)$.

\hspace*{\parindent}\textit{Proof}:  Note that with the proper variable partitioning in our proposed algorithm,  there is no constraint coupling between the variables in different blocks, as seen from $\left( {{\rm{\bar P}}1{\rm{ - }}3} \right)$, $\left( {{\rm{\bar P}}1{\rm{ - }}4} \right)$, and $\left( {{\rm{\bar P}}1{\rm{ - }}6} \right)$. 
 In addition, in  step $4$ of Algorithm~\ref{alg2},   $\left( {{\rm{\bar P}}1{\rm{ - }}3} \right)$ is solved by   an SCA method  and  a locally optimal solution is obtained. In  step $5$,   a globally optimal solution is obtained by using the Lagrange duality method for  $\left( {{\rm{\bar P}}1{\rm{ - }}4} \right)$. In  step $6$,   a locally optimal solution is obtained by using the MM method to solve $\left( {{\rm{\bar P}}1{\rm{ - }}6} \right)$. Following Theorem 4.1 in \cite{shi2016joint} together with the fact that for each subproblem in the inner layer   at least a locally optimal solution is obtained,  the proposed algorithm  is guaranteed to find  a locally optimal solution of $\left( {{\rm{\bar P}}1} \right)$. 
%
%

\section{Bisection Search-based Algorithm for PSR Optimization  Problem}
In this section, we study the PSR scenario and minimize the BER of the IRS symbols. It can be readily seen that  the objective function of $\left( {{\rm{P}}2} \right)$ is   a monotonically decreasing function in   ${\left| {{{\bf{v}}^H}{\rm{diag}}\left( {{\bf{h}}_r^H} \right){\bf{Gw}}} \right|^2}$. Thus, we can equivalently maximize the corresponding  SNR instead. The problem can be recast as follows
\begin{align}
\left( {{\rm{\bar P}}2} \right)&\mathop {\max }\limits_{{\bf{w}},{\bf{v}},\beta }~ \beta \notag\\
&{\rm s.t.}~ \frac{{{{\left| {{{\bf{v}}^H}{\rm{diag}}\left( {{\bf{h}}_r^H} \right){\bf{Gw}}} \right|}^2}}}{{{\sigma ^2}}} \ge \beta ,\label{psrP2const2}\\
& \qquad  \eqref{csrP1const2}, \eqref{csrP1const3},\eqref{psrP2const1}.
\end{align} 
In the following, we   propose an efficient bisection search based algorithm to solve $\left( {{\rm{\bar P}}2} \right)$. However, the search range for  $\beta$ is in principle  infinite, which would  make the   proposed  algorithm inefficient. To tackle this issue,  we first  confine the search space by deriving    an  upper bound for  $\beta$. 
\subsection{Confined Search Range  }
%
%
Problem $\left( {{\rm{\bar P}}2} \right)$ can be solved by finding   the maximum value of $\beta$ that  satisfies all the constraints.  Based on \eqref{csrP1const2} and \eqref{psrP2const2}, we have the following inequality
\begin{align}
\beta  \le \frac{{  {{\left| {{{\bf{v}}^H}{\rm{diag}}\left( {{\bf{h}}_r^H} \right){\bf{Gw}}} \right|}^2}}}{{{\sigma ^2}}} \overset{(a)}{\le} \frac{{{P_{\max }}  {{\left\| {{{\bf{v}}^H}{\rm{diag}}\left( {{\bf{h}}_r^H} \right){\bf{G}}} \right\|}^2}}}{{{\sigma ^2}}},
\end{align}
where $(a)$ holds since  the optimal beamforming vector $\bf w$ that maximizes  ${{{\left| {{{\bf{v}}^H}{\rm{diag}}\left( {{\bf{h}}_r^H} \right){\bf{Gw}}} \right|}^2}}$ is  ${\bf{w}} = \frac{{\sqrt {{P_{\max }}} {{\bf{G}}^H}{\rm{diag}}\left( {{{\bf{h}}_r}} \right){\bf{v}}}}{{\left\| {{{\bf{G}}^H}{\rm{diag}}\left( {{{\bf{h}}_r}} \right){\bf{v}}} \right\|}}$.  Similar to Lemma 1, a surrogate function for $\frac{{{{\bf{v}}^H}{\rm{diag}}\left( {{\bf{h}}_r^H} \right){\bf{G}}{{\bf{G}}^H}{\rm{diag}}\left( {{{\bf{h}}_r}} \right){\bf{v}}}}{{{\sigma ^2}}}$ at the initial point ${\bf v}^r$  by  using MM method is given by 
\begin{align}
\hat g\left( {{\bf{v}}|{{\bf{v}}^r}} \right) = {\hat \lambda _{\max }}M - 2{\mathop{\rm Re}\nolimits} \left\{ {{{\bf{v}}^H}\left( {{\hat\lambda _{\max }}{{\bf{I}}_M} - {\bf{\hat A}}} \right){{\bf{v}}^r}} \right\} + {{\bf{v}}^{r,H}}\left( {{\hat\lambda _{\max }}{{\bf{I}}_M} - {\bf{\hat A}}} \right){{\bf{v}}^r}, \label{psr_f_surrgogate}
\end{align}
where ${\bf{\hat A}}{\rm{ = }}\frac{{{\rm{diag}}\left( {{\bf{h}}_r^H} \right){\bf{G}}{{\bf{G}}^H}{\rm{diag}}\left( {{{\bf{h}}_r}} \right)}}{{{\sigma ^2}}}$, and ${\hat\lambda _{\max }}$ represents  the maximum  eigenvalue of ${\bf{\hat A}}$.  As a result, an upper bound of  $\beta$ can be obtained by  solving  the following optimization problem
\begin{align}
\left( {{\rm{\bar P}}2{\rm{ - }}1} \right)&{\kern 1pt} {\kern 1pt} {\kern 1pt}  \mathop {\max }\limits_{\bf{v}} ~{P_{\max }} \hat g\left( {{\bf{v}}{\rm{|}}{{\bf{v}}^r}} \right)\notag\\
&{\rm{s}}{\rm{.t}}{\rm{.}}~ \eqref{csrP1const3}.
\end{align}
Obviously, in the $r$th iteration, the optimal solution of  problem $\left( {{\rm{\bar P}}2{\rm{ - }}1} \right)$, denoted by ${{\bf{v}}^{r+1}}$,   is given by ${{\bf{v}}^{{\rm{r+1}}}} = {\rm{ - }}\exp \left( {j\arg \left( {\left( {{\lambda _{\max }}{{\bf{I}}_M} - {\bf{\hat A}}} \right){{\bf{v}}^r}} \right)} \right)$.  We then successively update the IRS phase-shift  vector  ${{\bf{v}}^{r+1}}$  according to $\left( {{\rm{\bar P}}2{\rm{ - }}1} \right)$, until   convergence is achieved.  The converged objective value is denoted by ${\beta}^{\rm up}$.

For any fixed $\beta$, we have to check whether the following  problem $\left( {{\rm{\bar P}}2{\rm{ - }}2} \right)$   is feasible 
\begin{align}
\left( {{\rm{\bar P}}2{\rm{ - }}2} \right)&{\kern 1pt} {\kern 1pt} {\kern 1pt} {\rm{Find:}}{\kern 1pt} {\kern 1pt} {\kern 1pt} {\kern 1pt} {\kern 1pt} {\kern 1pt} {\kern 1pt} {\kern 1pt} {\kern 1pt} {\kern 1pt} \left\{ {{\bf{w}},{\bf{v}}} \right\}\notag\\
& {\rm{s}}{\rm{.t}}{\rm{.}}~\eqref{csrP1const2}, \eqref{csrP1const3},\eqref{psrP2const1},\eqref{psrP2const2}.
\end{align}
If  problem $\left( {{\rm{\bar P}}2{\rm{ - }}2} \right)$ is feasible,  this indicates that $\beta$ is a feasible solution of  problem $\left( {{\rm{\bar P}}2} \right)$ and   $\beta$ can be enlarged to pursue a higher objective value; otherwise  $\beta$ is  infeasible, which indicates that  $\beta$ is too  large.
However, problem $\left( {{\rm{\bar P}}2{\rm{ - }}2} \right)$ has no objective function. To make it more tractable, $\left( {{\rm{\bar P}}2{\rm{ - }}2} \right)$ can be transformed to
\begin{align}
\left( {{\rm{\bar P}}2{\rm{ - 3}}} \right)&\mathop {\max }\limits_{{\bf{w}},{\bf{v}}} {\left| {\frac{{{\bf{h}}_d^H{\bf{w}}}}{\sigma }} \right|^2} - \left( {{2^{{R_{psr,{\rm{th}}}}}} - 1} \right)\left( {\frac{{\rho {{\left| {{{\bf{v}}^H}{\rm{diag}}\left( {{\bf{h}}_r^H} \right){\bf{Gw}}} \right|}^2}}}{{{\sigma ^2}}} + 1} \right)\notag\\
& {\rm{s}}{\rm{.t}}{\rm{.}}~\eqref{csrP1const2}, \eqref{csrP1const3},\eqref{psrP2const2}.
\end{align}
The objective function of $\left( {{\rm{\bar P}}2{\rm{ - 3}}} \right) $ is obtained by performing simple  algebraic operations on \eqref{psrP2const1}. If  the  obtained objective value of $\left( {{\rm{\bar P}}2{\rm{ - 3}}} \right)$ is no smaller than zero at the optimal point $\{\bf w,\bf v\}$, this  indicates that problem $\left( {{\rm{\bar P}}2{\rm{ - 2}}} \right) $ is feasible; otherwise it is   no feasible.  It is observed that the optimization  variables in the objective function and constraint \eqref{psrP2const2} are  intricately coupled, which   motivates us to apply the block coordinate descent method to solve $\left( {{\rm{\bar P}}2{\rm{ - 3}}} \right) $ by properly partitioning
the optimization variables into different blocks. Specifically,  $\left( {{\rm{\bar P}}2{\rm{ - 3}}} \right) $ is divided into two subproblems, namely, the BS beamforming optimization  subproblem  and IRS phase shift optimization  subproblem, and then we alternately optimize the two subproblems until  convergence is reached.
\subsection{Lagrange Duality  Method for BS Beamforming Optimizaiton}
For any given  phase shift vector $\bf v$, the BS beamforming optimization subproblem is given by 
\begin{align}
\left( {{\rm{\bar P}}2{\rm{ - 4}}} \right)&\mathop {\max }\limits_{\bf{w}} {\left| {\frac{{{\bf{h}}_d^H{\bf{w}}}}{\sigma }} \right|^2} - \left( {{2^{{R_{psr,{\rm{th}}}}}} - 1} \right)\left( {\frac{{\rho {{\left| {{{\bf{v}}^H}{\rm{diag}}\left( {{\bf{h}}_r^H} \right){\bf{Gw}}} \right|}^2}}}{{{\sigma ^2}}} + 1} \right)\notag\\
& {\rm{s}}{\rm{.t}}{\rm{.}}~\eqref{csrP1const2}, \eqref{psrP2const2}.
\end{align}
Problem $\left( {{\rm{\bar P}}2{\rm{ - 4}}} \right)$ is still non-convex due to the non-convex objective function as well as non-convex constraint  \eqref{psrP2const2}. Note that both  the objective function and the left-hand-side of \eqref{psrP2const2} are quadratic functions, the SCA method can be applied to address  this difficulty  efficiently. Specifically, based on the first-order Taylor expansion at  any given point ${\bf w}^r$, we have the following inequality 
\begin{align}
{\left| {\frac{{{\bf{h}}_d^H{\bf{w}}}}{\sigma }} \right|^2} \ge  - {\left| {\frac{{{\bf{h}}_d^H{{\bf{w}}^r}}}{\sigma }} \right|^2} + \frac{{2{\mathop{\rm Re}\nolimits} \left\{ {{{\bf{w}}^{r,H}}{{\bf{h}}_d}{\bf{h}}_d^H{\bf{w}}} \right\}}}{{{\sigma ^2}}}\overset{\triangle}{=}f_1^{{\rm{lb}}}\left( {\bf{w}} \right),
\end{align}
\begin{align}
\frac{{ {{\left| {{{\bf{v}}^H}{\rm{diag}}\left( {{\bf{h}}_r^H} \right){\bf{Gw}}} \right|}^2}}}{{{\sigma ^2}}} &\ge  - \frac{{ {{\left| {{{\bf{v}}^H}{\rm{diag}}\left( {{\bf{h}}_r^H} \right){\bf{G}}{{\bf{w}}^r}} \right|}^2}}}{{{\sigma ^2}}} + \frac{{2 {\mathop{\rm Re}\nolimits} \left\{ {{{\bf{w}}^{r,H}}{{\bf{G}}^H}{\rm{diag}}\left( {{{\bf{h}}_r}} \right){\bf{v}}{{\bf{v}}^H}{\rm{diag}}\left( {{\bf{h}}_r^H} \right){\bf{Gw}}} \right\}}}{{{\sigma ^2}}}\notag\\
&\overset{\triangle}{=}f_2^{{\rm{lb}}}\left( {\bf{w}} \right).
\end{align}
It can be readily seen that both $f_1^{{\rm{lb}}}\left( {\bf{w}} \right)$ and $f_2^{{\rm{lb}}}\left( {\bf{w}} \right)$ are  linear and thus convex  w.r.t. ${\bf w}$.
As a result, for a given  point ${\bf w}^r$, we have the following optimization problem 
\begin{align}
\left( {{\rm{\bar P}}2{\rm{ - 5}}} \right)&\mathop {\max }\limits_{\bf{w}} f_1^{{\rm{lb}}}\left( {\bf{w}} \right) - \left( {{2^{{R_{psr,{\rm{th}}}}}} - 1} \right)\left( {\frac{{\rho {{\left| {{{\bf{v}}^H}{\rm{diag}}\left( {{\bf{h}}_r^H} \right){\bf{Gw}}} \right|}^2}}}{{{\sigma ^2}}} + 1} \right)\label{psr_beamfroming_const0}\\
&{\rm{s}}{\rm{.t}}{\rm{.}}{\kern 1pt} {\kern 1pt} {\kern 1pt} {\kern 1pt} {\kern 1pt} f_2^{{\rm{lb}}}\left( {\bf{w}} \right) \ge \beta, \label{psr_beamfroming_const1}\\
&\qquad \eqref{csrP1const2}.
\end{align}
Although $\left( {{\rm{\bar P}}2{\rm{ - 5}}} \right)$ is a convex optimization  problem and  can be solved with the  interior point method, the resulting computational complexity  is ${\cal O}\left( {{N^{3.5}}} \right)$. In the following, we exploit the  Lagrange duality  method to reduce the  complexity.
 Note that compared to  $\left( {{\rm{\bar P}}1{\rm{ - 4}}} \right)$ for   CSR in Section III-B, $\left( {{\rm{\bar P}}2{\rm{ - 5}}} \right) $ for  PSR has  a different objective function and an additional constraint in \eqref{psr_beamfroming_const1}.
  Define by ${\tau _1} \ge 0$   the dual variable associated with \eqref{csrP1const2}. The partial Lagrange function of $\left( {{\rm{\bar P}}2{\rm{ - 5}}} \right)$ is given by 
\begin{align}
{{\cal L}_1}\left( {{\bf{w}},{\tau _1}} \right)  &=  f_1^{{\rm{lb}}}\left( {\bf{w}} \right) - \left( {{2^{{R_{psr,{\rm{th}}}}}} - 1} \right)\left( {\frac{{\rho {{\left| {{{\bf{v}}^H}{\rm{diag}}\left( {{\bf{h}}_r^H} \right){\bf{Gw}}} \right|}^2}}}{{{\sigma ^2}}} + 1} \right)\notag\\
& + {\tau _1}\left( {{P_{\max }} - \left\| {\bf{w}} \right\|_2^2} \right).
\end{align}
Thus, the corresponding dual function is given as follow 
\begin{align}
\left( {{\rm{\bar P}}2{\rm{-5-dual}}} \right)&{\kern 1pt} \mathop {\max }\limits_{\bf{w}} {\cal L}_1\left( {{\bf{w}},{\tau _1}} \right)\notag\\
&{\rm{s}}{\rm{.t}}{\rm{.}}{\kern 1pt} {\kern 1pt} {\kern 1pt} {\kern 1pt} {\kern 1pt} \eqref{psr_beamfroming_const1}.
\end{align}
To maximize $\left( {{\rm{\bar P}}2{\rm{-5-dual}}} \right)$ for a  given $\tau_1$, we introduce the  dual variable ${\tau _2} \ge 0$  associated with \eqref{psr_beamfroming_const1}. Then,  the 
Lagrange function of $\left( {{\rm{\bar P}}2{\rm{ - 5 - dual}}} \right)$ is given as follow 
\begin{align}
{{\cal L}_2}\left( {{\bf{w}},{\tau _2}} \right) = {{\cal L}_1}\left( {{\bf{w}},{\tau _1}} \right){\rm{ + }}{\tau _2}\left( {f_2^{{\rm{lb}}}\left( {\bf{w}} \right){\rm{ - }}\beta } \right).
\end{align}
By taking  the first-order derivative of ${{\cal L}_2}\left( {{\bf{w}},{\tau _2}} \right)$ w.r.t. $\bf w$  and setting it to zero, we  obtain
\begin{align}
{\bf{w}}\left( {{\tau _1},{\tau _2}} \right){\rm{ = }}{\left( {{{\bf{D}}_1}{\rm{ + }}{\tau _1}{\bf{I}}} \right)^\dag }\left( {\frac{{{{\bf{h}}_d}{\bf{h}}_d^H{{\bf{w}}^r}}}{{{\sigma ^2}}}{\rm{ + }}{\tau _2}{{\bf{d}}_2}} \right), \label{psr_beamforming}
\end{align}
where ${{\bf{D}}_1}{\rm{ = }}\left( {{2^{{R_{psr,{\rm{th}}}}}} - 1} \right)\left( {\frac{{\rho {{\left( {{{\bf{v}}^H}{\rm{diag}}\left( {{\bf{h}}_r^H} \right){\bf{G}}} \right)}^H}{{\bf{v}}^H}{\rm{diag}}\left( {{\bf{h}}_r^H} \right){\bf{G}}}}{{{\sigma ^2}}}} \right)$ and ${{\bf{d}}_2}{\rm{ = }}\frac{{\rho {{\bf{G}}^H}{\rm{diag}}\left( {{{\bf{h}}_r}} \right){\bf{v}}{{\bf{v}}^H}{\rm{diag}}\left( {{\bf{h}}_r^H} \right){\bf{G}}{{\bf{w}}^r}}}{{{\sigma ^2}}}$. For any  given $\tau_1$, the optimal value of $\tau_2$ must be chosen such that  the following complementary slackness condition is satisfied:
\begin{align}
\tau _2^{{\rm{opt}}}\left( {\beta {\rm{ - }}f_2^{{\rm{lb}}}\left( {{\bf{w}}\left( {{\tau _1},\tau _2^{{\rm{opt}}}} \right)} \right)} \right){\rm{ = }}0.
\end{align}
As such, if $\beta {\rm{ - }}f_2^{{\rm{lb}}}\left( {{\bf{w}}\left( {{\tau _1},0} \right)} \right) < 0$ holds, the optimal beamformer is ${{\bf{w}}\left( {{\tau _1},0} \right)}$; otherwise, the optimal beamformer is  ${{\bf{w}}\left( {{\tau _1},\tau _2^{{\rm{opt}}}} \right)}$  with ${\tau _2^{{\rm{opt}}}}$ given by 
\begin{align}
{\tau^{\rm opt} _2}{\rm{ = }}\frac{{\beta {\rm{ + }}\frac{{\rho {{\left| {{{\bf{v}}^H}{\rm{diag}}\left( {{\bf{h}}_r^H} \right){\bf{G}}{{\bf{w}}^r}} \right|}^2}}}{{{\sigma ^2}}}{\rm{ - }}2{\mathop{\rm Re}\nolimits} \left\{ {{\bf{d}}_2^H{{\left( {{{\bf{D}}_1}{\rm{ + }}{\tau _1}{\bf{I}}} \right)}^{\dag}}\frac{{{{\bf{h}}_d}{\bf{h}}_d^H{{\bf{w}}^r}}}{{{\sigma ^2}}}} \right\}}}{{2{\mathop{\rm Re}\nolimits} \left\{ {{\bf{d}}_2^H{{\left( {{{\bf{D}}_1}{\rm{ + }}{\tau _1}{\bf{I}}} \right)}^{\dag}}{{\bf{d}}_2}} \right\}}}. \label{psr_dual_tau_2}
\end{align}
To solve $\left( {{\rm{\bar P}}2{\rm{-5}}} \right)$, we need to calculate the optimal $\tau^{\rm opt}_1$. The  optimal value of $\tau_1$ must be chosen to  satisfy  the following complementary slackness condition
\begin{align}
\tau _1^{{\rm{opt}}}\left( {\left\| {{\bf{w}}\left( {\tau _1^{{\rm{opt}}},{\tau^{\rm opt} _2}} \right)} \right\|_2^2{\rm{ - }}{P_{\max }}} \right){\rm{ = }}0.
\end{align}
If ${\left\| {{\bf{w}}\left( {0,\tau _2^{{\rm{opt}}}} \right)} \right\|^2} \le {P_{\max }}$ holds,
the optimal beamformer is given by ${\bf{w}}\left( {0,\tau _2^{{\rm{opt}}}} \right)$; otherwise, we need to calculate the optimal $\tau^{\rm opt}_1$ that satisfies ${P_{\max }}{\rm{ = }}\left\| {{\bf{w}}\left( {\tau _1^{{\rm{opt}}},\tau _2^{{\rm{opt}}}} \right)} \right\|_2^2$. However, since there is no closed-form expression for  $\tau_1$ w.r.t. $\tau_2$, it is difficult to show  that  $\left\| {{\bf{w}}\left( {\tau _1,\tau _2} \right)} \right\|_2^2$  is  monotonic w.r.t. $\tau_1$. This problem is addressed in the following lemma. 

\textbf{\emph{Lemma 3:}} $\left\| {{\bf{w}}\left( {\tau _1,\tau _2} \right)} \right\|_2^2$ is a non-increasing function of   $\tau_1$.

\hspace*{\parindent}\textit{Proof}: Please refer to the Appendix~\ref{appendix1}.

Based on Lemma 3, we can use  a  bisection search-based method to find the optimal $\tau_1$. The details of the proposed Lagrange duality  method for solving $\left( {{\rm{\bar P}}2{\rm{-5}}} \right)$  are summarized in Algorithm~\ref{alg3}.
\begin{algorithm}[!t]
	\caption{ Lagrange  duality  method  for solving problem $\left( {{\rm{\bar P}}2{\rm{-5}}} \right)$.}
	\label{alg3}
	\begin{algorithmic}[1]
		\STATE  \textbf{Initialize} ${\tau_1 ^{{\rm{lb}}}}$, ${\tau_1 ^{{\rm{up}}}}={{\bar \tau}_1 ^{{\rm{up}}}}$, and  $\varepsilon$.
		 \STATE If  $\beta {\rm{ - }}f_2^{{\rm{lb}}}\left( {{\bf{w}}\left( {0,0} \right)} \right) \le 0$ holds, $\tau_1^{\rm opt}=0$ and  $\tau_2^{\rm opt}=0$; otherwise $\tau_1^{\rm opt}=0$, and $\tau_2^{\rm opt}$ is given by \eqref{psr_dual_tau_2}.
		 \STATE  If $\left\| {{\bf{w}}\left( {0,\tau _2^{{\rm{opt}}}} \right)} \right\|_2^2 \le {P_{\max }}$, the optimal beamformer is given by ${\bf{w}}\left( {0,\tau _2^{{\rm{opt}}}} \right)$, and go to step 10; otherwise go to step 4.
		\STATE \textbf{Repeat}
		\STATE \quad Compute $\tau_1  = \frac{{{\tau_1 ^{{\rm{lb}}}} + {\tau_1 ^{{\rm{up}}}}}}{2}$.
		\STATE  \quad If  $\beta {\rm{ - }}f_2^{{\rm{lb}}}\left( {{\bf{w}}\left( {{\tau _1},0} \right)} \right) \le 0$ holds, $\tau_2^{\rm opt}$ is set to zero; otherwise $\tau_2^{\rm opt}$ is given by \eqref{psr_dual_tau_2}. 
		\STATE \quad Update beamformer ${\bf{w}}\left( {{\tau _1},{\tau _2^{\rm opt}}} \right)$ according to \eqref{psr_beamforming}.
			
		\STATE \quad If $\left\| {{\bf{w}}\left( {{\tau _1},\tau _2^{{\rm{opt}}}} \right)} \right\|_2^2 \le {P_{\max }}$, set ${\tau_1 ^{{\rm{up}}}} = \tau_1 $, otherwise, set ${\tau_1 ^{{\rm{lb}}}} = \tau_1 $.
		\STATE \textbf{Until} $\left| {{\tau_1 ^{{\rm{up}}}} - {\tau_1 ^{{\rm{lb}}}}} \right| \le \varepsilon $.
		\STATE If $\left| {{\tau_1 ^{{\rm{lb}}}} - {{\bar \tau_1} ^{{\rm{up}}}}} \right| \le \varepsilon $, which indicates problem $\left( {{\rm{\bar P}}2{\rm{ - 5}}} \right)$ is infeasible, we terminate the algorithm.
		\STATE {\bf Output:}  Optimal BS beamformer ${{\bf{w}}\left( {\tau _1^{{\rm{opt}}},\tau _2^{{\rm{opt}}}} \right)}$ according to \eqref{psr_beamforming}.
	\end{algorithmic}
\end{algorithm}

\subsection{DC Method for IRS Phase Shift Optimization}
For any given BS beamformer $\bf w$, by ignoring  constants  that do not depend on  $\bf v$, the IRS phase shift optimization subproblem can be  formulated as  follows
\begin{align}
\left( {{\rm{\bar P}}2{\rm{ - 6}}} \right)&\mathop {\min }\limits_{\bf{v}} \frac{{{{\left| {{{\bf{v}}^H}{\rm{diag}}\left( {{\bf{h}}_r^H} \right){\bf{Gw}}} \right|}^2}}}{{{\sigma ^2}}}\notag\\
&{\rm s.t.}~\eqref{csrP1const3},\eqref{psrP2const2}.
\end{align}
Due to the unit-modulus constraint in \eqref{csrP1const3}, a commonly used approach  is to reformulate $\left( {{\rm{P}}2{\rm{ - 6}}} \right)$ as a semidefinite
programming (SDP) problem \cite{wu2019intelligent,wu2019weighted}.  Specifically,  define ${\bf{V}}{\rm{ = }}{\bf{v}}{{\bf{v}}^H}$, which needs to satisfy ${\bf V}\succeq {\bf 0}$ and ${\rm{rank}}\left( {\bf{V}} \right) = 1$. As a result, problem $\left( {{\rm{\bar P}}2{\rm{ - 6}}} \right)$ is equivalent  to
\begin{align}
\left( {{\rm{\bar P}}2{\rm{ - 7}}} \right)&{\kern 1pt} {\kern 1pt} {\kern 1pt} \mathop {\min }\limits_{\bf{V}\succeq {\bf 0}} {\rm{tr}}\left( {{\bf{VB}}} \right)\notag\\
& {\rm{s}}{\rm{.t}}{\rm{.}}{\kern 1pt} {\kern 1pt} {\kern 1pt} {\kern 1pt} {\kern 1pt} {\rm{tr}}\left( {{\bf{VB}}} \right) \ge \beta,\label{psr_p2_7_const1}\\
&\qquad {{\bf{V}}_{i,i}} = 1,\forall m,\label{psr_p2_7_const2}\\
&\qquad {\rm{rank}}\left( {\bf{V}} \right) = 1, \label{psr_p2_7_const3}
\end{align}
where ${\bf{B}}{\rm{ = diag}}\left( {{\bf{h}}_r^H} \right){\bf{Gw}}{\left( {{\rm{diag}}\left( {{\bf{h}}_r^H} \right){\bf{Gw}}} \right)^H}/{\sigma ^2}$. It can be seen that the objective function, and constraints \eqref{psr_p2_7_const1}  and \eqref{psr_p2_7_const2} are all linear w.r.t. $\bf V$, while constraint \eqref{psr_p2_7_const3} is non-convex.
A common method for addressing this issue  is to apply SDR     by dropping the non-convex rank-one constraint, i.e., constraint \eqref{psr_p2_7_const3}, and then solve the relaxed problem    via    standard convex optimization techniques \cite{wu2019intelligent}. If the  solution $\bf V$ of  the relaxed  version of problem  $\left( {{\rm{\bar P}}2{\rm{ - 7}}} \right)$ is rank-one, the optimal  phase shift $\bf v$ can be optimally obtained by applying  Cholesky decomposition of  $\bf V$. Otherwise,  the  Gaussian randomization technique
can be applied  to construct a rank-one solution from the obtained
high-rank solution  $\bf V$\cite{sidiropoulos2006transmit}. However,   Gaussian randomization   may not be able to   guarantee a locally and/or globally optimal solution, especially 
when  the dimension of matrix $\bf V$ (which is equal to the number of IRS reflecting elements) is large. To overcome this drawback,  we apply   DC programming  to solve $\left( {{\rm{P}}2{\rm{ - 7}}} \right)$, which  guarantees  convergence to a KKT point \cite{tao1997convex},\cite{Yang2020Federated}. We first introduce the following important lemma needed for the development of the proposed DC method.
 
\textbf{\emph{Lemma 4:}}  For a  positive semidefinite  matrix $\bf V$ and ${\rm{rank}}\left( {\bf{V}} \right) \ge 1$,  we have the following equivalence  \cite{Yang2020Federated}, \cite{jiang2019over},
\begin{align}
{\rm{rank}}\left( {\bf{V}} \right) = 1 \Leftrightarrow {\rm{tr}}\left( {\bf{V}} \right) - {\left\| {\bf{V}} \right\|_2} = 0.
\end{align}
Note that it can be readily checked that  ${{\rm{tr}}\left( {\bf{V}} \right) \ge {{\left\| {\bf{V}} \right\|}_2}}$. By adding the term ${{\rm{tr}}\left( {\bf{V}} \right) - {{\left\| {\bf{V}} \right\|}_2}}$ in the objective function of  $\left( {{\rm{\bar P}}2{\rm{ - 7}}} \right)$ as a penalized term, problem $\left( {{\rm{\bar P}}2{\rm{ - 7}}} \right)$ can be  rewritten as follows
\begin{align}
\left( {{\rm{\bar P}}2{\rm{ - 8}}} \right)&{\kern 1pt} {\kern 1pt} {\kern 1pt} \mathop {\min }\limits_{{\bf{V}} \succeq {\bf{0}}} {\rm{tr}}\left( {{\bf{VB}}} \right) + \frac{1}{\bar \eta} \left( {{\rm{tr}}\left( {\bf{V}} \right) - {{\left\| {\bf{V}} \right\|}_2}} \right)\notag\\
&{\rm s.t.}~\eqref{psr_p2_7_const1},\eqref{psr_p2_7_const2},
\end{align}
where ${\bar \eta }$ is a penalty  coefficient.   Then,  we can  apply   a similar  two-stage penalty-based method to solve $\left( {{\rm{\bar P}}2{\rm{ - 8}}} \right)$ as was presented in Section III. Specifically, we update the penalty coefficient ${\bar \eta }$  in the outer layer, and  solve the penalized optimization problem in the inner layer. In the inner layer, for a  fixed ${\bar \eta }$, the objective function of $\left( {{\rm{\bar P}}2{\rm{ - 8}}} \right)$ is not convex and  is still difficult to solve. The main idea behind   DC programming  is to construct a sequence of convex  surrogates  to replace the non-convex term, and solve the  constructed convex surrogates in an iterative manner. Specifically, by linearizing the  term $-{{{\left\| {\bf{V}} \right\|}_2}}$ at a given point ${\bf V}^r$ at the $r$th iteration, we obtain  the following optimization problem 
\begin{align}
\left( {{\rm{\bar P}}2{\rm{ - 9}}} \right)&{\kern 1pt} {\kern 1pt} {\kern 1pt} \mathop {\min }\limits_{{\bf{V}}\succeq {\bf 0}} {\rm{tr}}\left( {{\bf{VB}}} \right) + \frac{1}{\bar \eta}  {\rm{Re}}\left\{ {{\rm{tr}}\left( {\left( {{\bf{I}}{\rm{ - }}{{\left( {\partial {{\left\| {{{\bf{V}}^r}} \right\|}_2}} \right)}^H}} \right){\bf{V}}} \right)} \right\}\notag\\
&{\rm s.t.}~\eqref{psr_p2_7_const1},\eqref{psr_p2_7_const2},
\end{align}
where ${\partial {{\left\| {{{\bf{V}}^r}} \right\|}_2}}$ denotes the subgradient of $\bf V$ at  point ${\bf V}^r$. It is worth pointing out that  
 ${\partial {{\left\| {{{\bf{V}}^r}} \right\|}_2}}$ can be calculated  from Proposition 4 of  reference \cite{Yang2020Federated} and  is given by 
\begin{align}
\partial {\left\| {\bf{V}} \right\|_2}{\rm{ = }}{\bf{v}}_p{{\bf{v}}_p^H},
\end{align}
where ${{\bf{v}}_p}$ denotes the eigenvector corresponding to the largest eigenvalue of ${\bf V}$. Both the  objective function and the constraints of $\left( {{\rm{\bar P}}2{\rm{ - 9}}} \right)$ are convex.   Thus, $\left( {{\rm{\bar P}}2{\rm{ - 9}}} \right)$ can be efficiently solved by the standard convex optimization techniques \cite{boyd2004convex}.  We then successively update ${\bf V}$ obtained from $\left( {{\rm{\bar P}}2{\rm{ - 9}}} \right)$, until   convergence is reached.  Note that  the  solution $\bf V$ obtained from $\left( {{\rm{\bar P}}2{\rm{ - 9}}} \right)$ after convergence must be rank-one, we thus can uniquely reconstruct the beamforming vector $\bf v$ from the obtained solution  $\bf V$ via  Cholesky
decomposition.
\subsection{Overall Algorithm}
Based on the solutions to  the above subproblems, a bisection search-based  method  is proposed, which is summarized in Algorithm~\ref{alg4}. Note that if the finally obtained  objective value of $\left( {{\rm{\bar P}}2{\rm{ - 3}}} \right)$, denoted by ${f}_{\rm obj}$, is smaller than zero, this  indicates that there is no feasible solution for problem $\left( {{\rm{\bar P}}2} \right)$. Since   stationary points are obtained for both blocks defined by problems $\left( {{\rm{\bar P}}2{\rm{ - 5}}} \right)$ and $\left( {{\rm{\bar P}}2{\rm{ - 9}}} \right)$ in steps 6 and 7, respectively,   Algorithm~\ref{alg4} is  guaranteed to converge to a KKT solution of problem $\left( {{\rm{\bar P}}2} \right)$\cite{yu2020irs}. The     computational complexity of Algorithm~\ref{alg4} can be determined  as follows. In step $2$, the complexity of computing  ${{\beta ^{{\rm{up}}}}}$ is mainly caused by the calculation of    the maximum  eigenvalue of  ${\bf{\hat A}}$, which is given by ${\cal O}\left( {{M^3}} \right)$.
In step $6$, the complexity of calculating  beamforming vector  $\bf w$  via the  Lagrange duality  method is ${\cal O}\left( {{{\log }_2}\left( {\frac{{\tau _1^{{\rm{up}}} - \tau _1^{{\rm{lb}}}}}{\varepsilon }} \right){N^3}} \right)$. In step $7$,  the complexity of  calculating the  IRS phase shift matrix   $\bf V$ based on  SDP is ${\cal O}\left( {{M^{3.5}}} \right)$. Therefore, the overall  complexity of Algorithm~\ref{alg4} is given by  ${\cal O}\left( {{M^3} + {{\log }_2}\left( {\frac{{{\beta ^{{\rm{up}}}} - {\beta ^{{\rm{lb}}}}}}{{{\varepsilon _2}}}} \right)\left( {{{\log }_2}\left( {\frac{{\tau _1^{{\rm{up}}} - \tau _1^{{\rm{lb}}}}}{\varepsilon }} \right){N^3} + I{M^{3.5}}} \right)} \right)$, where $I$ denotes the number of iterations required by the penalty-based method  for reaching convergence.

\begin{algorithm}[!t]
	\caption{Bisection search-based algorithm for solving problem $\left( {{\rm{\bar P}}2} \right)$.}
	\label{alg4}
	\begin{algorithmic}[1]
		\STATE  \textbf{Initialize}  $\beta^{\rm lb}$,  $\bar \eta$, 	 phase shift vector ${\bf v}^{r_1}$,  beamforming vector ${\bf w}^{r_1}$,  predefined thresholds $\varepsilon_1$,  $\varepsilon_2$, iteration index  $r_1=0$.
		\STATE  Calculate $\beta^{\rm up}$ based on $\left( {{\rm{\bar P}}2{\rm{ - 1}}} \right)$.
		\STATE  \textbf{Repeat}
	   \STATE \quad Calculate $\beta  = \frac{{{\beta ^{{\rm{lb}}}} + {\beta ^{{\rm{up}}}}}}{2}$.
		\STATE  \quad \textbf{Repeat}
	    \STATE \qquad Update ${\bf w }^{r_1+1}$ in  problem $\left( {{\rm{\bar P}}2{\rm{ - 5}}} \right)$ by using Algorithm~\ref{alg3}. If problem $\left( {{\rm{\bar P}}2{\rm{ - 5}}} \right)$ is \\
	    \qquad  infeasible, set ${\beta ^{{\rm{up}}}} = \beta $ and go to step $4$.
        \STATE \qquad Update ${\bf v}^{r_1+1}$ in  $\left( {{\rm{\bar P}}2{\rm{ - 9}}} \right)$ by using a penalty-based method as in  Algorithm~\ref{alg2}.
	    \STATE \quad \textbf{Until}  the fractional increase of the  objective value of $\left( {{\rm{\bar P}}2{\rm{ - 3}}} \right)$ is below  $\varepsilon_1$.
	    \STATE  \quad Calculate  the objective value of $\left( {{\rm{\bar P}}2{\rm{ - 3}}} \right)$, denoted by ${f}_{\rm obj}$. If the value of ${f}_{\rm obj}\ge0$, set \\
	    \quad ${\beta ^{{\rm{lb}}}} = \beta $,  $r_1=0$; otherwise, set ${\beta ^{{\rm{up}}}} = \beta $, $r_1=0$.
	    \STATE  \textbf{Until} $\left| {{\beta ^{{\rm{up}}}} - {\beta ^{{\rm{lb}}}}} \right| \le {\varepsilon _2}$.	
	\end{algorithmic}
\end{algorithm}

\section{Numerical Results}
In this section, we provide numerical results  to validate  the performance of the proposed algorithms  for   IRS-based SR  transmission systems. We assume that the BS is equipped with    a uniform linear array with  $N=10$ elements, while the IRS  is equipped with  a uniform rectangular array with $M=M_xM_z$, where $M_x$ and $M_z$ denote the numbers of reflecting elements along the $x$-axis and $z$-axis, respectively. We fix $M_x=5$ and increase $M_z$ linearly with $M$.  We assume that the antenna spacing is half a  wavelength. The BS, IRS, and IR are  located at $( 0, 0,0)$, $(100 ~\rm m, 0 , 2.5 ~\rm m)$, and $(100 ~\rm m, 0,0)$ in   $3$D  Cartesian coordinates, respectively. In addition, the large-scale path loss is modeled as   ${L_{\rm loss}} = {{L}_0}{\left( {{d \over {{d_0}}}} \right)^{ - \alpha }}$, where ${ L}_0$ denotes the channel power gain at the reference distance of  ${{d_0}}=1~\rm m$,  $d$ is the link distance, and $\alpha$ is the path loss exponent. 
We assume that  the  BS-IRS and  IRS-IR links are     Rician fading with a Rician factor of $3~\rm dB$, and the  BS-IR link is     Rayleigh fading. In addition, the path loss exponents for the BS-IRS, IRS-IR, and BS-IR links are set as $2.6$, $2.6$, and $3.6$, respectively.
   Unless otherwise stated, we set  $R_{csr,{\rm th}}=1~ {\rm bps/Hz}$, $R_{psr,{\rm th}}=1~ {\rm bps/Hz}$, ${{{L}}_0} =  - 30~{\rm{dB}}$, $P_{\max}=40~\rm dBm$, $\eta=0.1$, ${\bar \eta}=10^2$, ${\sigma ^2} =  - 80~{\rm{dBm}}$, $\rho=0.5$, $L=15$, $c=0.7$,  $\varepsilon=10^{-6}$,  $\varepsilon_1=\varepsilon_2=10^{-4}$,  $\beta^{\rm lb}=0$, ${\tau_1 ^{{\rm{lb}}}}=\beta^{\rm lb}=0$, ${\lambda ^{{\rm{lb}}}}=10^{-5}$, and ${\tau_1 ^{{\rm{up}}}}={{\bar \tau}_1 ^{{\rm{up}}}}=10^6$.

\begin{figure*}[!t]
	\centering
	\subfigure[Constraint violation $\xi$.]{
		\begin{minipage}[t]{0.5\linewidth}
			\centering
			\includegraphics[width=2.6in]{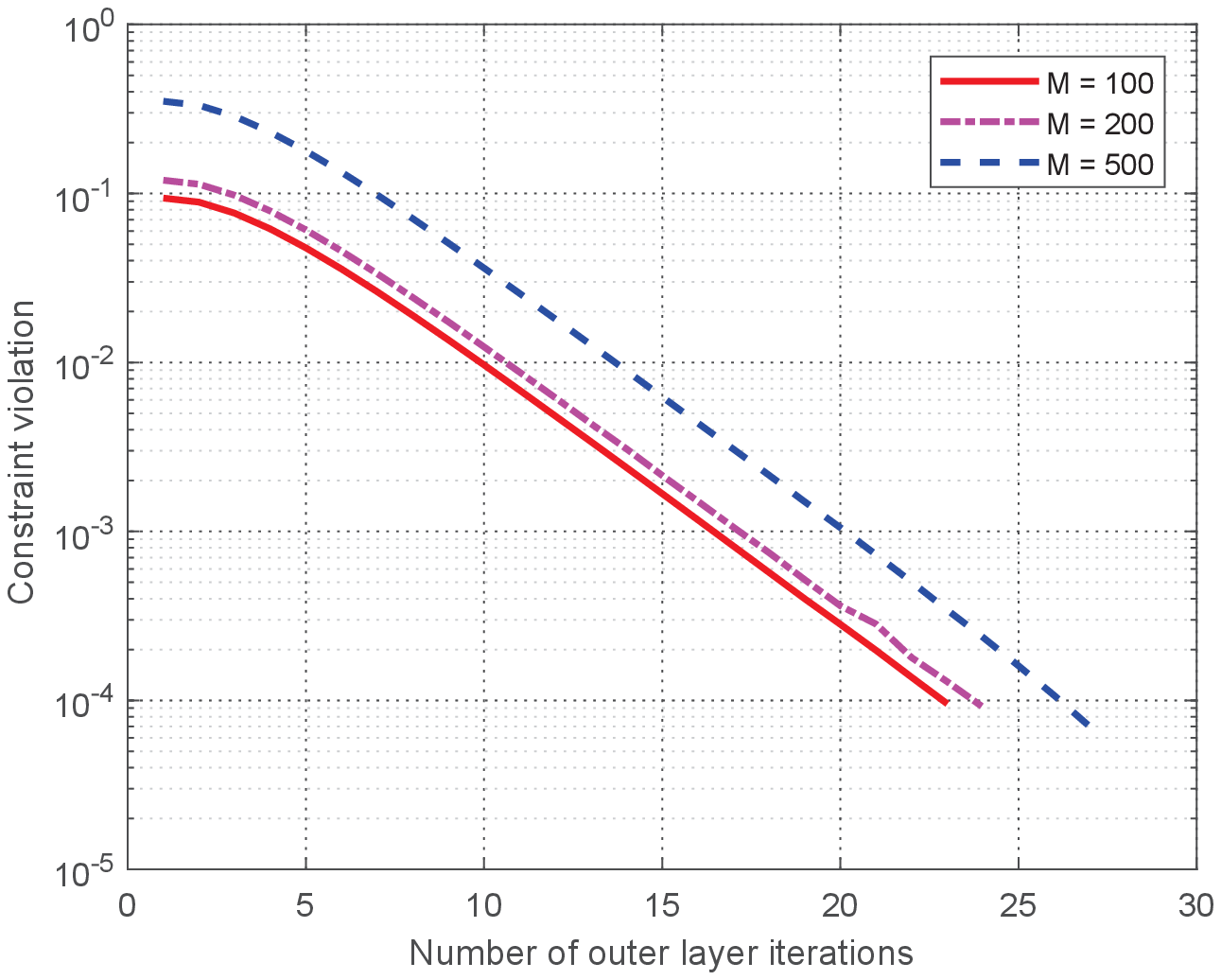}
		\end{minipage}%
	}
	\subfigure[The penalized objective value.]{
		\begin{minipage}[t]{0.5\linewidth}
			\centering
			\includegraphics[width=2.6in]{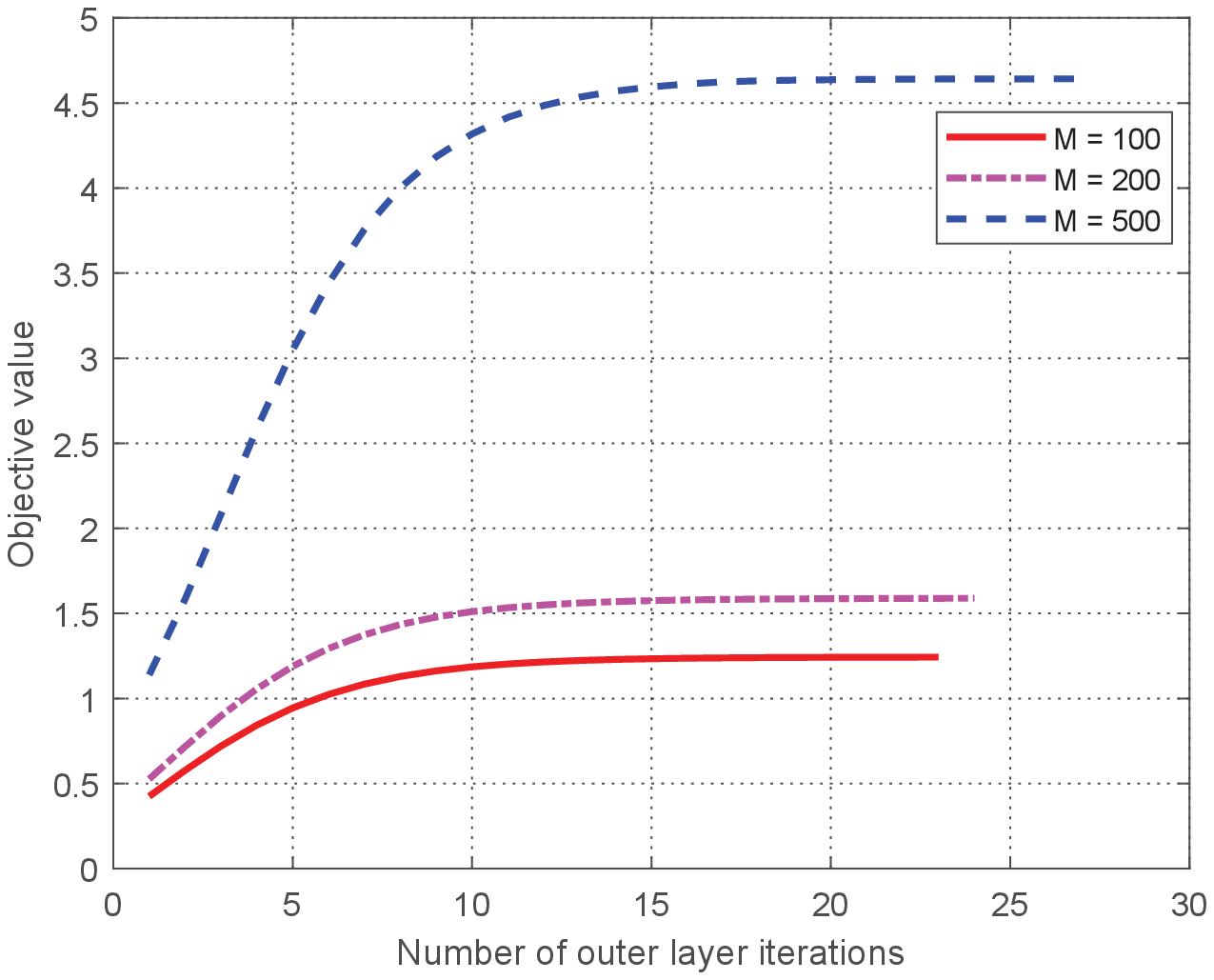}
		\end{minipage}%
	}%
	\quad             
	\centering
	\caption{One channel realization is considered to illustrate  the  convergence behaviour of Algorithm~\ref{alg2}.}\label{csr_fig1}
		\vspace{-15pt}
\end{figure*} 
Before discussing the  performance  of  the  proposed schemes, we  first  verify  the effectiveness of the proposed penalty-based Algorithm~\ref{alg2} for  CSR. The constraint violation and convergence behaviour of Algorithm~\ref{alg2} are shown  in Fig.~\ref{csr_fig1} for  one channel realization  for     different numbers of IRS reflecting elements $M$, namely, $M=100$, $M=200$, and $M=500$. From Fig.~\ref{csr_fig1}(a), it  is observed that the  constraint violation $\xi$  converges very fast  to the  predefined violation accuracy of  $10^{-4}$ after about  $23$ iterations for $M=100$, which indicates that   equality constraints  \eqref{csrbarP1const2} and \eqref{csrbarP1const3}  in $\left( {{\rm{\bar P}}1} \right)$ are eventually satisfied. Even for $M=500$, only $27$ iterations are required for reaching the  predefined violation accuracy, which demonstrates the  effectiveness of Algorithm~\ref{alg2}. This can   be observed more clearly in Fig.~\ref{csr_fig1}(b), where 
 the penalized objective values of $\left( {{\rm{\bar P}}1{\rm{ - }}1} \right)$ obtained for   different $M$ all  increase quickly with the number of iterations and finally converge.

\begin{figure}[!t]
	\centering
	\begin{minipage}[t]{0.45\textwidth}
		\centering
		\includegraphics[width=2.7in]{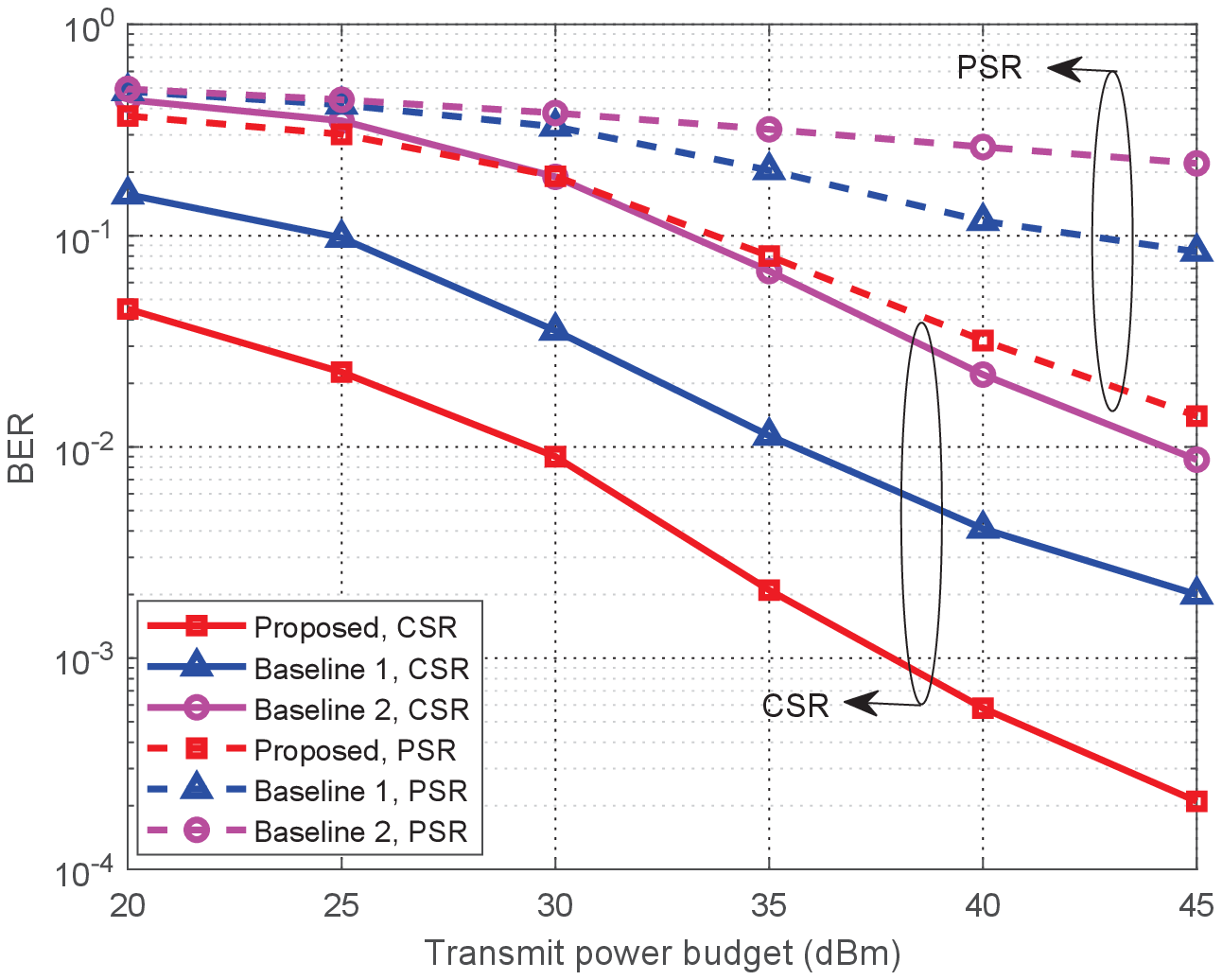}
		\caption{BER versus  transmit power budget  $P_{\max}$.}\label{csr_fig2}
	\end{minipage}
	\begin{minipage}[t]{0.49\textwidth}
		\centering
		\includegraphics[width=2.7in]{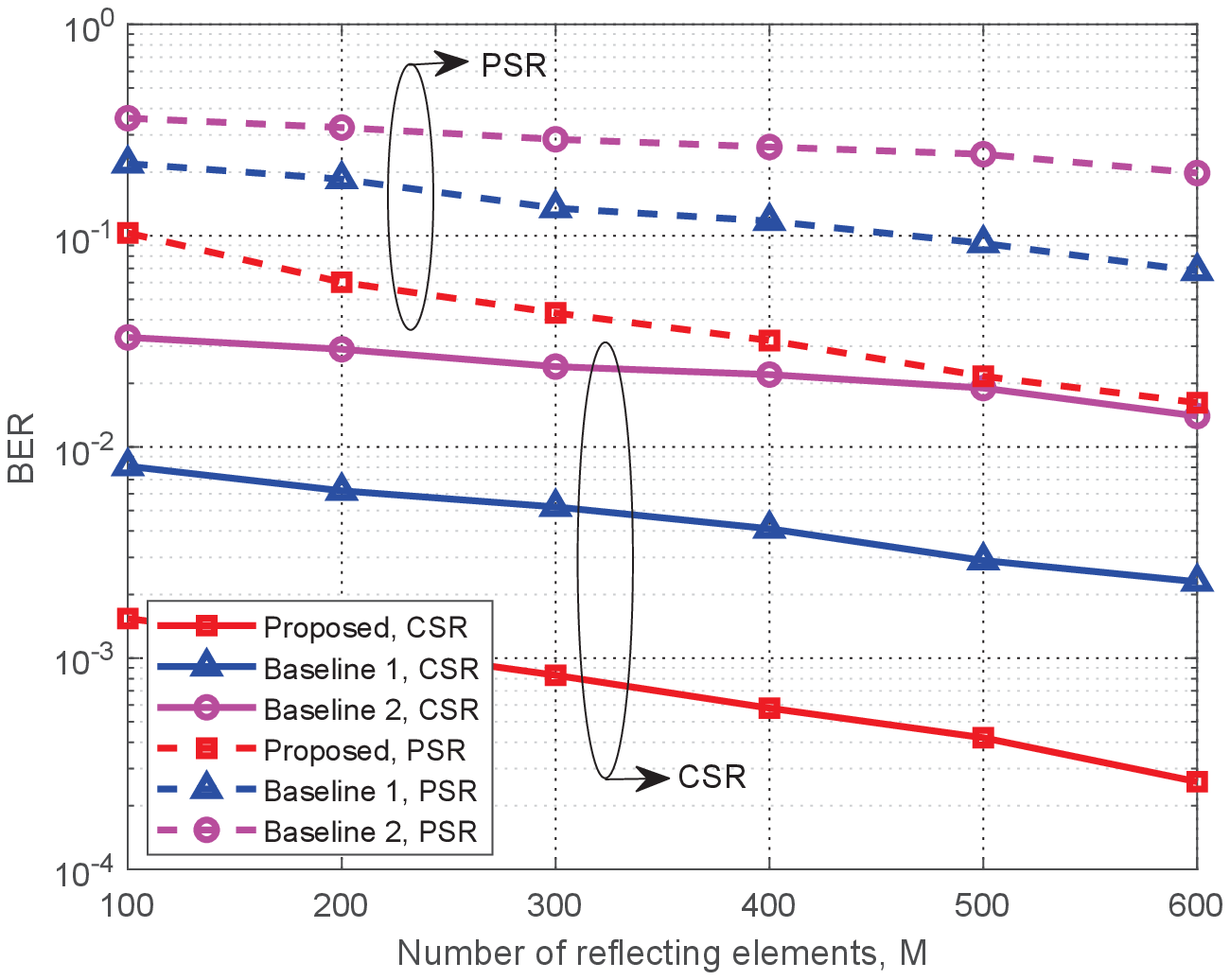}
		\caption{BER versus  the number of IRS reflecting elements $M$.}\label{csr_fig3} 
	\end{minipage}
\vspace{-15pt}
\end{figure}

In order to evaluate the performance  of the proposed   IRS-based SR   system, we compare  the following  schemes for CSR  and PSR:  
1) Proposed scheme: We jointly optimize the BS beamformer and phase shifts to minimize the BER of the IRS symbols. For CSR,   Algorithm~\ref{alg2} is used, while for PSR,   Algorithm~\ref{alg4} is applied; 2) Baseline Scheme $1$: we set ${\bf{w}} = {{\sqrt {{P_{\max }}} {{\bf{h}}_d}} \mathord{\left/
		{\vphantom {{\sqrt {{P_{\max }}} {{\bf{h}}_d}} {\left\| {{{\bf{h}}_d}} \right\|}}} \right.
		\kern-\nulldelimiterspace} {\left\| {{{\bf{h}}_d}} \right\|}}$ to achieve MRT for the BS-IR direct link, and the BER of the IRS symbols is minimized by optimizing the phase shifts; and 3) Baseline Scheme $2$: the  IRS phase shifts are random and follow  uniform distributions, and  the BER of the IRS symbols is minimized by optimizing the BS beamformer. 
In Fig.~\ref{csr_fig2}, we compare  the BER of the IRS symbols obtained for  the above schemes versus  $P_{\max}$ for   $M=400$. Note that all   results shown are obtained by simulation where we   average $200$ channel realizations. As can be  observed for all  considered schemes,    the BER of the IRS symbols    decreases with $P_{\max}$. This is expected  since   from the objective functions of $(\rm P1)$ and $(\rm P2)$,  it can be easily seen that  as $P_{\max}$ grows, the SNR   increases   with the transmit power, thereby reducing  the BER of the IRS symbols. In addition, for CSR, it is observed that  the proposed  scheme outperforms both  Baseline Schemes $1$  and $2$,   which demonstrates that the BER of the IRS symbols can  indeed be reduced significantly with the joint BS  beamformer and IRS phase shift  optimization.  A similar  behavior  is also observed for   PSR.
 Furthermore, it is observed that the BER of the IRS symbols for  CSR is   significantly lower than that for PSR.  This is because for  CSR, one IRS symbols spans $L$ primary symbols. Thus, a diversity gain is obtained  by exploiting    MRT to coherently add up the multi-path signals to increase the SNR at  the  receiver. In contrast,  for  PSR,  the period of the   IRS symbol is equal to that of  the primary  symbol. Hence, the IRS reflected signals are treated as interference for the primary network, which thus degrades the  system performance. In other words,  the interference is harnessed in  CSR, while it  is harmful  in PSR.

In Fig.~\ref{csr_fig3},  we show the BER of the IRS symbols   versus the number of reflecting elements $M$. It is observed that  the BER of the IRS symbols obtained by  all considered  schemes   decreases  with $M$.  This is because   more  reflecting elements help achieve a higher passive beamforming gain, thereby improving the SNR. More importantly,   since the IRS is passive with  low power consumption and low hardware  cost, it is promising to apply large IRSs with  hundreds   of  reflecting elements. Moreover, for CSR, our proposed scheme outperforms Baseline Scheme $1$, which illustrates the benefits  introduced by    BS beamformer optimization. 
Furthermore,    Baseline Scheme $2$   achieves some  performance gains for CSR since   the IRS is able to reflect some of the dissipated signals back to the receiver.  A similar  behavior  is also observed for   PSR. In addition, similar to  Fig.~\ref{csr_fig2}, the BER of the IRS symbols obtained with  CSR  is   lower than  that obtained with   PSR  since the interference is harnessed in   CSR. 

\begin{figure}[!t]
	\centering
	\begin{minipage}[t]{0.45\textwidth}
		\centering
		\includegraphics[width=2.7in]{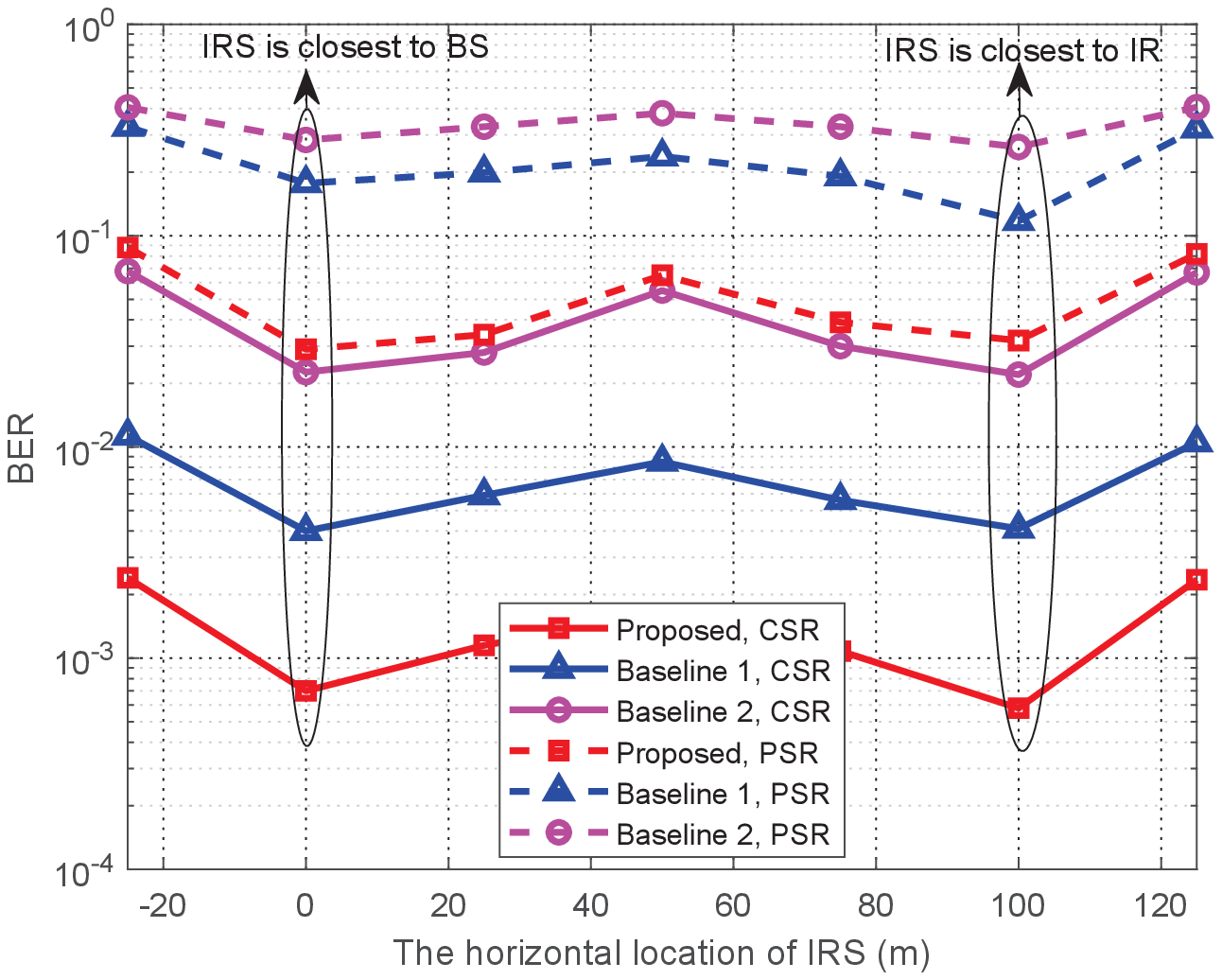}
		\caption{The impact of IRS location on  BER.}\label{csr_fig4}
	\end{minipage}
	\begin{minipage}[t]{0.45\textwidth}
		\centering
		\includegraphics[width=2.7in]{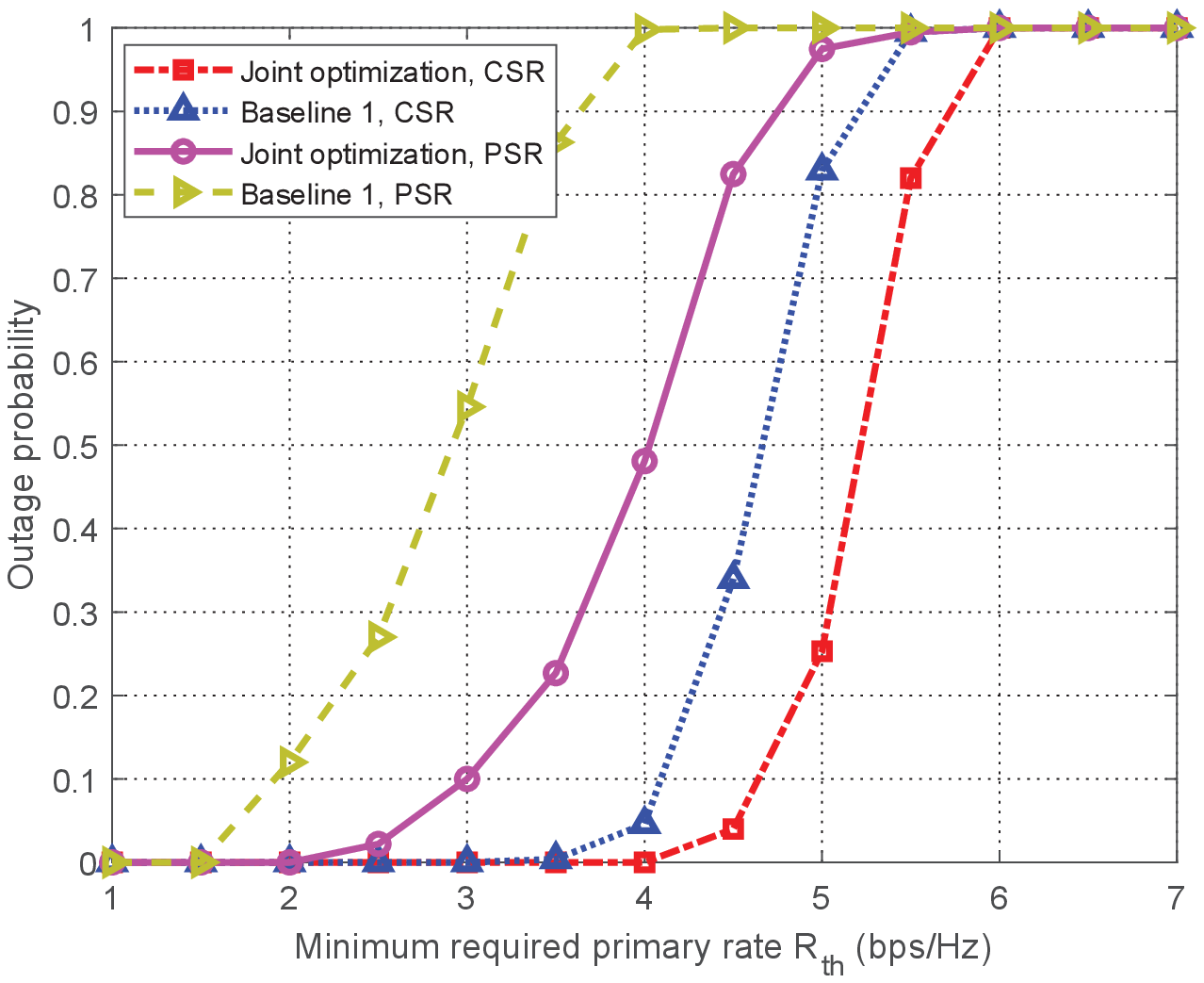}
		\caption{Outage probability versus   $R_{\rm th}$.}\label{csr_fig5} 
	\end{minipage}
	\vspace{-15pt}
\end{figure}

In Fig.~\ref{csr_fig4}, we  study the  impact of the  IRS location on the    BER  of the IRS symbols for  $M=400$.  Specifically, we study the BERs  obtained with the considered schemes versus the IRS's horizontal location ($x$-coordinate), ranging from $-25~\rm m$ to $125~\rm m$. Note that for $x=0~\rm m$,  the IRS is  closest to the BS, while for  $x=100~\rm m$, the IRS is  closest to the IR.
As can be  observed,  if  the IRS is deployed close  to the BS  or IR, the   BER decreases.  This is because for a short distance between   IRS and BS or IR, the signal attenuation in the   BS-IRS-IR link is reduced due to the smaller double path loss \cite{wu2020intelligentarxiv}. Additionally, for CSR, the proposed  scheme  still outperforms   Baseline Schemes $1$  and $2$. Similar results are also obtained for   PSR.   This further  demonstrates the benefits of the proposed  joint IRS phase shift  and BS beamforming optimization.

\begin{figure}[!t]
 	\centerline{\includegraphics[width=2.7in]{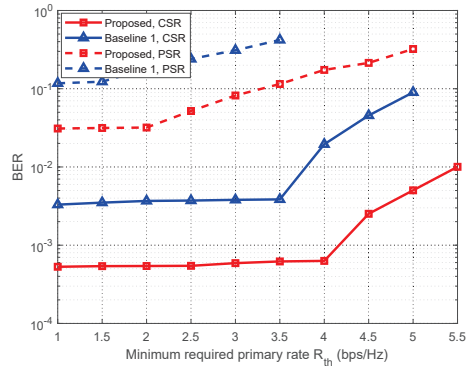}}
 	\caption{BER versus   $R_{\rm th}$.} \label{fig6}
 		\vspace{-15pt}
 \end{figure}

In Fig.~\ref{csr_fig5}, we  plot the outage probability of  the proposed schemes versus the required   primary  rate. For ease of exposition, we set $R_{csr,\rm th}=R_{psr,\rm th}=R_{\rm th}$. The outage probability is defined as the probability that the received primary rate  at the IR  is lower than a predefined minimum required primary rate  $R_{\rm th}$. For the joint optimization scheme for CSR, we check  the feasibility of $(\rm P1)$ by jointly optimizing  the beamformer and the  phase shifts, while for Baseline Scheme $1$,  we   optimize the  phase shifts for the given   MRT beamformer for the BS-IR link. For PSR,  we make a similar comparisons for $(\rm P2)$.
It is observed that the outage probability of all  schemes increases  with  $R_{\rm th}$ and  approaches $1$ for large $R_{\rm th}$. This is expected since the primary rate is upper bounded by a finite value due to the limited  BS transmit power budget. For the CSR scenario,  the joint optimization scheme has a lower  outage probability than    Baseline  Scheme $1$, especially when  $R_{\rm th}$ is larger than    $4~{\rm bps/Hz}$. For example, the outage probability for the joint optimization scheme is about  $0.042$ for  $R_{\rm th}=4.5~{\rm bps/Hz}$,  while that for   Baseline  Scheme $1$  is about  $0.339$. This    can also  be deduced from   $\left( {{\rm{ P}}1} \right)$. For any given phase shift $\bf v$, the    left-hand-side of \eqref{csrP1const1} obtained by optimizing the BS beamformer $\bf w$ is larger than that  obtained by applying MRT, which indicates that the  joint optimization scheme has a higher probability of satisfying constraint  \eqref{csrP1const1}. For the PSR scenario,  the joint optimization  scheme has a lower  outage probability than    Baseline  Scheme $1$. This   can be  deduced from $\left( {{\rm{ P}}2} \right)$.  For any given phase shifts $\bf v$, the   left-hand-side of \eqref{psrP2const1} obtained by optimizing the BS beamformer $\bf w$ is larger than that  obtained by applying MRT. In addition, the outage probability of the joint optimization scheme for CSR is lower than that for PSR. This can be readily derived  from \eqref{csrP1const1} and \eqref{psrP2const1}, where the   left-hand-side of \eqref{csrP1const1} is evidently larger than that of \eqref{psrP2const1}, which implies that  a higher primary rate can be obtained with  CSR. 
To see this  more clearly, Fig.~\ref{fig6} studies the BER of the IRS symbols  versus $R_{\rm th}$. It can be observed that for small $R_{\rm th}$,  the BER  remains nearly unchanged, while for large $R_{\rm th}$,  the BER increases substantially. This is because as $R_{\rm th}$ becomes larger,  the primary rate requirement constraint becomes stringent, so the  optimization of  the  beamformer and phase shifts needs to   fulfill the primary rate requirement at the cost of  sacrificing the system performance. 

\section{Conclusion}
In this paper, we have studied  novel paradigms for  IRS-based SR systems. Depending on the IRS’s  symbol period,  two scenarios, namely, CSR and PSR, have been considered with the objective of minimizing  the BER of the IRS symbols by jointly optimizing the active beamformer at the  base station and  the  phase shifts at the IRS while guaranteeing  the minimum primary rate requirements.  For the CSR scenario, we have decomposed the original problem into three  subproblems, which  allowed us to
obtain  semi-closed-form solutions for the  BS beamformer   and  the IRS phase shifts. Then,   a penalty-based algorithm with a two-stage iteration  has been proposed to obtain a high-quality solution. For the PSR scenario,   a bisection search based  algorithm has been proposed. In particular, we have obtained a semi-closed-form solution for the BS beamformer, and  leveraged the DC  programming  framework  to obtain a  rank-one solution.  Our simulation results have shown that  the  proposed SR techniques  achieve lower BERs as compared with   two benchmark schemes and demonstrated  that the BER  can be   significantly reduced by jointly optimizing the BS beamformer and   IRS phase shifts for both scenarios. In addition, our results  have also shown that the BER can be significantly reduced by   
 proper  positioning of the IRS, especially by placing   the IRS   close to the BS and/or IR. The results in this paper can be further extended by considering multiple IRSs, frequency-selective channel models, imperfect CSI, etc., which are interesting topics   for future work in this area.

\appendices
\section{Proof of Lemma~3} \label{appendix1}
Define   two dual variables $\tau_1$ and $\tau _1^{'}$ corresponding  to problem $\left( {{\rm{\bar P}}2{\rm{ - 5}}} \right)$.  Then, the corresponding beamforming vectors are denoted by ${{\bf{w}}\left( {{\tau _1},\tau _2^{{\rm{opt}}}} \right)}$ and ${{\bf{w}}\left( {\tau _1^{'},\tau _2^{{\rm{{'}opt}}}} \right)}$, respectively. In addition, we   set $\tau_1>\tau _1^{'}$. Since ${{\bf{w}}\left( {{\tau _1},\tau _2^{{\rm{opt}}}} \right)}$ is the optimal beamformer with  given ${{\tau _1}}$, we   have 
\begin{align}
&{{\cal L}_1}\left( {{\bf{w}}\left( {{\tau _1},\tau _2^{{\rm{opt}}}} \right),{\tau _1}} \right) \ge {{\cal L}_1}\left( {{\bf{w}}\left( {\tau _1^{'},\tau _2^{{\rm{{'}opt}}}} \right),{\tau _1}} \right),\\
&{{\cal L}_1}\left( {{\bf{w}}\left( {\tau _1^{'},\tau _2^{{\rm{{'}opt}}}} \right),\tau _1^{'}} \right) \ge {{\cal L}_1}\left( {{\bf{w}}\left( {\tau ,\tau _2^{{\rm{opt}}}} \right),\tau _1^{'}} \right).
\end{align}
Adding the above two inequalities, we have  
\begin{align}
\left( {{\tau _1} - \tau _1^{'}} \right)\left( {{{\left\| {{\bf{w}}\left( {\tau _1^{'},\tau _2^{{\rm{{'}opt}}}} \right)} \right\|}^2} - {{\left\| {{\bf{w}}\left( {{\tau _1},\tau _2^{{\rm{opt}}}} \right)} \right\|}^2}} \right) \ge 0.
\end{align}
Since $\tau_1>\tau _1^{'}$, we directly arrive at  ${\left\| {{\bf{w}}\left( {\tau _1^{'},\tau _2^{{\rm{{'}opt}}}} \right)} \right\|^2} \ge {\left\| {{\bf{w}}\left( {{\tau _1},\tau _2^{{\rm{opt}}}} \right)} \right\|^2}$. This thus completes the proof. 
\bibliographystyle{IEEEtran}
\bibliography{IRS_symbiotic}
\end{document}